\documentclass[a4paper,fleqn,usenatbib]{mnras}

\usepackage[T1]{fontenc}
\usepackage{ae,aecompl}
\usepackage[utf8]{inputenc}

\usepackage{graphicx}	
\usepackage{amsmath}	
\usepackage{amssymb}	
\usepackage{lscape}
\usepackage{tikz}   
\usepackage{tikz-3dplot} 
\usepackage{import}
\usepackage{lmodern}
\RequirePackage{fix-cm}

\usepackage{graphicx}	
\usepackage{amsmath}	
\usepackage{amssymb}	
\usepackage{grffile}

\newcommand{\rlight}{r_{\rm L}}

\newcommand{\rot}{\mathbf{\nabla} \times}
\newcommand{\divg}{\mathbf{\nabla}\cdot}

\DeclareRobustCommand{\rchi}{{\mathpalette\irchi\relax}}
\newcommand{\irchi}[2]{\raisebox{\depth}{$#1\chi$}} 

\title[Radiative pulsar magnetospheres]{Radiative pulsar magnetospheres: oblique rotators}

\author[J\'er\^ome P\'etri]{
J. P\'etri\thanks{E-mail: jerome.petri@astro.unistra.fr}
\\
Universit\'e de Strasbourg, CNRS, Observatoire astronomique de Strasbourg, UMR 7550, F-67000 Strasbourg, France.
}

\date{Accepted XXX. Received YYY; in original form ZZZ}

\pubyear{2020}

\begin{document}
\label{firstpage}
\pagerange{\pageref{firstpage}--\pageref{lastpage}}
\maketitle

\begin{abstract}
Pulsar magnetospheres are filled with relativistic pairs copiously emitting photons detected from the radio wavelengths up to high and very high energies, in the GeV and sometimes in the TeV range. Efficient particle acceleration converts the stellar rotational kinetic energy into radio, X-ray and gamma-ray photons. Force-free magnetospheres, being dissipationless, cannot operate this conversion. Some non ideal plasma effects must set in within the magnetosphere. In this paper, we compute numerical solutions of pulsar radiative magnetospheres in the radiation reaction limit, where radiation fully balances single particle acceleration. Using an appropriate Ohm's law, the dissipation is only controlled by the pair multiplicity factor~$\kappa$. Moreover we allow for either a minimal radiative region where dissipation is added only where required or for a force-free inside radiative outside model. This approach naturally and self-consistently connects the particle dynamics to its radiation field in the ultra-relativistic regime. Our solutions tend to the force-free limit for moderately large multiplicities, $\kappa \gg 1$, decreasing the spin-down energy conversion into radiation. Nevertheless, for sufficiently low multiplicity $\kappa \lesssim1$, a significant fraction of the spin-down energy flows into radiation via particle acceleration. The work done by the electromagnetic field on the plasma mainly occurs in the current sheet of the striped wind, right outside the light-cylinder. Nevertheless the impact on the magnetic topology is negligible whatever the model. Therefore the associated sky maps and light-curves are only weakly impacted as shown.
\end{abstract}




\section{Introduction}

Pulsars are very efficient particle accelerators as witnessed by they broad band electromagnetic spectrum from radio \citep{manchester_australia_2005} up to very high-energy, GeV \citep{abdo_second_2013}, and sometimes TeV emission like for the Crab \citep{ansoldi_teraelectronvolt_2016} and Vela \citep{djannati-atai_h.e.s.s._2017}. Particle acceleration and therefore radiation is rooted to the fast rotation of a strongly magnetized neutron star. Rotational kinetic energy is converted into radiation by curvature, synchrotron and inverse Compton emission leading to the stellar braking accounted by the spin-down rate derived from the period~$P$ and its derivative~$\dot{P}$. However, where exactly within the magnetosphere those mechanisms occur is still unclear. Undoubtedly particles flow at very high Lorentz factor from the star to the interstellar medium, shaping the pulsar wind as a ballerina similar to the solar wind. The global magnetosphere electrodynamics is intimately related to the motion of these particles and their subsequent radiation. Some localized regions are prone to efficient conversion of the rotational kinetic energy into acceleration and radiation but where and how remains to be self-consistently determined from global magnetosphere simulations including dissipation.

The simplest approach to find such solutions starts with force-free regime (FFE) where an ideal plasma is considered, neglecting particle inertia and temperature, meaning that the electric field $\mathbf{E}$ is orthogonal to the magnetic field $\mathbf{B}$, $\mathbf{E} \cdot \mathbf{B}=0$ and the electric field is weaker than the magnetic field in normalized units, meaning $E<c\,B$ where $c$ is the speed of light.
In this picture, the Poynting flux is conserved because the electromagnetic field does no work on the plasma via the electric current~$\mathbf{j}$, meaning $\mathbf{j} \cdot \mathbf{E}=0$.
Strictly speaking, such magnetospheres are invisible because no photons are produces. 

Numerical simulations have been pioneered by \cite{contopoulos_axisymmetric_1999} for the axisymmetric rotator that was extended to an oblique rotator by \cite{spitkovsky_time-dependent_2006} and retrieved by other authors, whether only aligned \citep{timokhin_force-free_2006, cao_spectral_2016, komissarov_simulations_2006, parfrey_introducing_2012, chen_electrodynamics_2014} or oblique \citep{petri_pulsar_2012, kalapotharakos_three-dimensional_2009, tchekhovskoy_three-dimensional_2016}. See also different solutions not requiring a current sheet like for instance in \cite{lovelace_jets_2006}. 

The aforementioned fluid description offers a good starting point to understand the global electric circuit made of charge and current densities. However, it neglects some fundamental kinetic aspects required to self-consistently include single particle acceleration as well as radiation feedback. As kinetic simulations are much more demanding than fluid models, this approach was only scarcely investigated in the last century. Let us mention \cite{krause-polstorff_electrosphere_1985} who computed axisymmetric dead pulsar magnetospheres called electrospheres. Due to the axisymmetry of the problem they used rings of charges instead of point particles. Later with the advent on computational power, \cite{smith_numerical_2001} showed with slightly more sophisticated simulations that a fully field magnetosphere is unstable and collapse to an electrosphere. The first full three-dimensional electrosphere was constructed by \cite{mcdonald_investigations_2009} using an electromagnetic Particle in Cell (PIC) code. They neglect pair creation and therefore did not add any particle injection process. Eventually \cite{philippov_ab_2014} computed the first two-dimensional axisymmetric pulsar magnetosphere for an aligned rotator by permanently injecting particle supposed to be released from the surface, avoiding to end to an electrosphere configuration \citep{petri_global_2002}. Depending on the volume injection rate, they were able to find any equilibrium between the force-free and the fully charge separated state. \cite{chen_electrodynamics_2014} improved this model by adding a prescription for the pair creation, putting a threshold on the lepton Lorentz factor. Following the same lines \cite{cerutti_particle_2015} assumed particle injection only from the vicinity of the stellar surface. \cite{belyaev_dissipation_2015} injected particles from regions where a parallel electric field exists. The first full three dimensional PIC simulations of a pulsar magnetosphere were performed by \cite{philippov_ab_2015}. For an aligned rotator \cite{philippov_ab_2015-1} also included general-relativistic corrections with frame-dragging. Soon after some observational signature predictions were added to compute light curves and spectra emanating from curvature and or synchrotron radiation like for instance \cite{cerutti_modelling_2016} who then included polarization \citep{cerutti_polarized_2016}. This PIC simulations were then extended to the striped wind well outside the light-cylinder to study its dissipation \citep{cerutti_dissipation_2017, cerutti_dissipation_2020}. The oblique magnetosphere with radiation and general-relativistic correction was eventually computed by \cite{philippov_ab-initio_2018}. Several other groups performed similar simulations like \cite{brambilla_electronpositron_2018} or \cite{kalapotharakos_fermi_2017} and \cite{kalapotharakos_three-dimensional_2018} who tried to explicitly connect their simulation results to gamma-ray observations. Alternatively, more simply test particle trajectories can be explored within a fluid code, see for instance \cite{brambilla_testing_2015}.

Even if the PIC approach is now mature to include several ingredients like pair creation and its subsequent radiative signature, its main flaw resides in its inability to simulate neutron star magnetospheres with realistic stellar magnetic field strengths and rotation periods. For instance the Larmor radius is 10 to 15 orders of magnitude smaller than the light-cylinder radius, putting stringent constraints on the time step than cannot be fulfilled with current computational technology. It is therefore difficult to connect straightforwardly the microphysics dynamics induced by the gyro motion to the dynamics of the global magnetosphere although the time and spatial scale hierarchy is maintained in current PIC simulations. Consequently, a fluid description, as the one we employ in this paper, remains a valuable tool to explore the dynamics of pulsar magnetospheres. A hybrid approach using a particle kinetic description wherever necessary and a fluid model elsewhere would represent a good compromise. Recent developments indeed combine the PIC technique to the MHD evolution as for instance performed by \cite{marle_magnetic_2018} for particle shock acceleration or by \cite{bai_magnetohydrodynamic-particle--cell_2015} for investigation of cosmic rays interaction with a thermal plasma. Such hybrid modules are also implemented in available MHD codes like PLUTO \citep{mignone_particle_2018}. This new trend highlights the need to pursue our ongoing effort on improving plasma fluid models jointly with particle and Vlasov techniques by adding more physics on macro and micro scales simultaneously.

Returning to a simpler fluid and not particle description of the magnetosphere, the next step requires a proper treatment of dissipation with radiation. Some resistive simulations have been performed but with a resistivity not always based on pure physical grounds or with some arbitrariness leading to no unique prescription \citep{li_resistive_2012, kalapotharakos_toward_2012, gruzinov_strong-field_2008}. Although the best view would be a full kinetic description including, acceleration and radiation is conceptually possible, we believe it better at this stage to use a fluid description of the plasma by deriving an Ohm's law according to the radiation reaction limit and derived in for instance \cite{mestel_stellar_1999}.
\cite{contopoulos_are_2016}, build on this work and found radiative magnetospheric solution for oblique rotators however in a simplified manner. Recently a fully radiative solution has been computed for aligned rotators by \cite{petri_radiative_2020} and extended to oblique rotators \cite{petri_electrodynamics_2020} assuming that force-free dynamics holds inside the light-cylinder. \cite{cao_three-dimensional_2020} found similar solutions but allowing also possible dissipation within the light cylinder. Recently, \cite{cao_pulsar_2022} computed high resolution radiative magnetosphere solutions including some test particle dynamics in order to predict synchrotron spectra and light curves.

In the radiation reaction limit, the equation of motion is solved for a single particle in a stationary regime where the Lorentz force is counterbalanced by a so called radiative friction for ultra-relativistic speeds. Assuming that the particle moves at exactly the speed of light leads to a unique solution for the velocity vector given some decades ago by \cite{mestel_stellar_1999}. This expression is sometimes also called Aristotelian electrodynamics. Electrons and positrons, although possessing the same electric drift motion because being independent of the particle charge, will move in opposite direction with respect to the perpendicular plane. Their charge density will generate a current leading to a one parameter family of current prescription reminded in the next section. Therefore, it should be clear that radiative simulations as those we present in this paper include single particle dynamics concretized through the electric current density prescription.

The radiation reaction approximation solves the single particle equation of motion in the ultra-relativistic regime where the speed is exactly equal to the speed of light. The solution to the Lorentz force leads to the Aristotelian dynamics, the velocity being only a function of the local electric and magnetic field. However in order to avoid complication due to finite masses, as in the force-free case, we neglect their inertia. In this picture, particles only contribute to the charge and current density required to solve Maxwell equations. The derived current density possesses a component along the electric field and leads naturally to a kind of resistivity.

A full description of the plasma in this limit requires knowledge of its lepton content, separating the contribution from the electrons and the positrons. This pair multiplicity factor~$\kappa$ serves this goal and is the only free parameter in this radiative Ohm law. A fully self-consistent picture must however add pair-production but this small scale physics is still difficult to reconcile with the global scale of the magnetosphere. No self-consistent simulations have been performed so far taking all the ingredients self-consistently into account. Nevertheless, non thermal acceleration and radiative feedback is included in the present work thanks to the Aristotelian dynamics.

Because the radiation reaction limit regime relies on assumptions note always met within the magnetosphere, it is worth keeping in mind several caveats of our approach. The velocity field in Aristotelian electrodynamics is derived in the limit of significant radiative friction in the Lorentz force, reaching on a short time scale an asymptotic regime of exact balance between electric field acceleration and radiation damping. While this regime could be achieved in many places within the magnetosphere, there exist localized regions where such intense radiation damping is not effective due to negligible radiation reaction. Indeed, in the vicinity of polar caps, pair production efficiently screens the electric field component parallel to the magnetic field and particles do not experience the radiation reaction force \citep{timokhin_current_2013}. Moreover radiation damping involves ultra-relativistic particles with very large Lorentz factors that fails to be produced at several places except maybe for a tiny population of highly energetic particles. The outcome is a complex particle distribution function resembling more to a power law than to a mono-energetic population we take in this paper. As will be shown, Aristotelian dynamics allows for large regions where the electric field~$E$ dominates the magnetic field~$B$, i.e. where $E > c\,B$. However, recent studies showed that this should only occur as a transition stage to a magnetically dominated regime where $E < c\,B$ \citep{li_fast_2021}, see also \cite{beskin_radio_2018}. Moreover, our treatment neglects magnetic reconnection, especially within the current sheet of the striped wind outside the light cylinder, although it has been observed in PIC and MHD simulations. Our scheme represents a simplified two stage process where dissipation is directly converted into radiation.

In this paper, we compute oblique pulsar magnetospheres in the radiation reaction limit taking into account the exact dissipative current containing the electric drift component as well as the components aligned with the electric and magnetic field. In Sec.~\ref{sec:Modele}, we describe the model of our radiative magnetosphere and the prescription for Ohm's law derived from the radiation reaction regime. Some examples of magnetic topologies for an aligned and an orthogonal rotator are presented in Sec.~\ref{sec:LigneChamp} for the ideal FFE field and for the radiative magnetospheres. Next, in Sec.~\ref{sec:Luminosite} we compute the spin-down luminosity extracted from these models and compare it with previous works. The importance of dissipation is pointed out in Sec.~\ref{sec:Dissipation}. The importance of the parallel electric field component is stressed in Sec.~\ref{sec:parallel_E} Influences on the polar cap shape and size is explored in Sec.~\ref{sec:Calottes}. Sky maps and light-curves are presented in Sec.~\ref{sec:emission}. Conclusions are drawn in Sec.~\ref{sec:Conclusion}.

\section{Magnetospheric model}
\label{sec:Modele}

In this section, we present the underlying model to compute radiative pulsar magnetospheres starting from Maxwell equations and then explaining the electric current prescription.

\subsection{Maxwell equations}

In our models, the plasma only furnishes the required charge~$\rho_{\rm e}$ and current~$\mathbf{j}$ densities to evolve Maxwell equations written in standard MKSA units as
\begin{subequations}
	\begin{align}
	\label{eq:Maxwell1}
	\divg \mathbf B & = 0 \\
	\label{eq:Maxwell2}
	\rot \mathbf E & = - \frac{\partial \mathbf B}{\partial t} \\
	\label{eq:Maxwell3}
	\divg \mathbf E & = \frac{\rho_{\rm e}}{\varepsilon_0} \\
	\label{eq:Maxwell4}
	\rot \mathbf B & = \mu_0 \, \mathbf j + \frac{1}{c^2} \, \frac{\partial \mathbf E}{\partial t}  .
	\end{align}
\end{subequations}
Apart from the obvious boundary conditions on the stellar surface, the current density $\mathbf{j}$ is the only unknown of the problem. Once fixed according to a given plasma model, Maxwell equations can be solved numerically, leading to a magnetosphere solution. So let us describe the possibilities for this current.

\subsection{Current prescription}

\subsubsection{Force-free limit}

The simplest model corresponds to an ideal plasma with infinite conductivity, leading to the force-free prescription as
\begin{equation}
	\label{eq:J_ideal}
	\mathbf j = \rho_{\rm e} \, \frac{\mathbf{E}\wedge \mathbf{B}}{B^2} + \frac{\mathbf{B} \cdot \rot \mathbf{B} / \mu_0 - \varepsilon_0 \, \mathbf{E} \cdot \rot \mathbf{E}}{B^2} \, \mathbf{B} .
\end{equation}
By construction, this current does not work on particles since $\mathbf{j} \cdot \mathbf{E} = 0$. All the rotational kinetic energy goes into the Poynting flux of the low frequency large amplitude electromagnetic wave. For completeness and comparison with other models shown in this paper, we compute again some force-free magnetospheres.

\subsubsection{Radiative solution}

If some emission is taken into account, for instance like in the radiation reaction limit of ultra-relativistic particles, the velocity of particles is fully determined by the local electromagnetic field configuration. Indeed, the friction caused by a radiative term can be seen as an isotropic emission of photons in the particle rest frame and at a rate controlled by the radiated power $\mathcal{P} \geq 0$ such that the balance between Lorentz force and radiative friction is
\begin{equation}\label{eq:balance_lorentz_radiation}
q \, ( \mathbf{E} + \mathbf{v} \wedge \mathbf{B}) = \frac{\mathcal{P}}{c^2} \, \mathbf{v} .
\end{equation}
A justification and argumentation about the validity of this balance can be found in \cite{mestel_axisymmetric_1985} starting from the Lorentz-Abraham-Dirac equation.
This equation is solved explicitly with respect to the velocity vector $\mathbf{v}$ and given for positive charges as $\mathbf{v}_+$ and negative charges as $\mathbf{v}_-$ according to 
\begin{equation}
\label{eq:VRR}
\mathbf{v}_\pm = \frac{\mathbf{E} \wedge \mathbf{B} \pm ( E_0 \, \mathbf{E} / c + c \, B_0 \, \mathbf{B})}{E_0^2/c^2+B^2} .
\end{equation}
This expression only assumes that particles move exactly at the speed of light.
$E_0$ and $B_0$ are the strength of the electric and magnetic field deduced from the electromagnetic invariants and satisfying $\mathcal{I}_1 = \bmath E^2 - c^2 \, \bmath B^2 = E_0^2 - c^2 \, B_0^2$ and $\mathcal{I}_2 = c \, \bmath E \cdot \bmath B = c\,E_0 \, B_0$. Explicitly solving for $E_0\geq0$ and $B_0$ we find
\begin{subequations}
	\label{eq:E0B0}
	\begin{align}
	E_0^2 & = \frac{1}{2} \, (\mathcal{I}_1 + \sqrt{\mathcal{I}_1^2 + 4 \, \mathcal{I}_2^2 }) \\
	c\,B_0 & = \textrm{sign} (\mathcal{I}_2) \, \sqrt{E_0^2 - \mathcal{I}_1} .
	\end{align}
\end{subequations}
$E_0$ and $B_0$ are interpreted as the electric and magnetic field strength in a frame where electric and magnetic field are parallel to each other.

The radiated power is then simply
\begin{equation}\label{eq:puissance_rayonnee}
\mathcal{P} = |q| \, E_0 \, c \geq 0 .
\end{equation}
Therefore within physical constants, $E_0$ is a direct measure of the radiated power.
Single particles are therefore evolved according to eq.~\eqref{eq:VRR}. It corresponds to the exact solution of the Lorentz equation of motion for charges subject to a friction and moving at exactly the speed of light. The particle inertia has been neglected and all species with the same sign of charge possess the same velocity vector irrespective of their charge to mass ratio~$q/m$ as long as their sign does not change.

From the velocity expression in \eqref{eq:VRR} we can derive the electric current associated to this particle flow. The detailed derivation is given by \cite{petri_theory_2016}, and the associated radiative current density~$\mathbf{j}$ with minimal assumption is explained in \cite{petri_radiative_2020}. The final expression reduces to
\begin{equation}
\label{eq:J_rad}
\mathbf j = \rho_{\rm e} \, \frac{\mathbf E \wedge \mathbf B}{E_0^2/c^2 + B^2} + ( |\rho_{\rm e}| \, + 2\,\kappa \, n_0  \, e) \, \frac{E_0 \, \mathbf E/c^2 + B_0 \, \mathbf B}{E_0^2/c^2 + B^2}
\end{equation}
where $\kappa$ is the pair multiplicity. 
The background particle density number is depicted by $n_0$ and varies from point to point within the magnetosphere. For a self-consistent picture, this density should be constrained by the pair production rate. However such task is out of the scope of the present study. In order to go further, we replace $n_0$ by $|\rho_{\rm e}/q|$ and therefore is directly connected to the electric field via Maxwell-Gauss equation. Consequently
\begin{equation}\label{eq:happa_n0}
|\rho_{\rm e}| \, + 2\,\kappa \, n_0  \, e = |\rho_{\rm e}| \, (1 + 2\,\kappa)
\end{equation}
but actually this factor could be any function in the most general situation.

We stress that in our simulations, individual particles follow the velocity given in eq.\eqref{eq:VRR}. Each particle possesses its own velocity depending solely on the local electromagnetic field configuration. Therefore particles are present indirectly in the simulations with an analytical expression for the velocity, the Aristotelian dynamics. Nevertheless we do not follow individual particle trajectory because this would also require some assumption about the particle injection rate and its spatial dependence. To avoid such arbitrariness, we fixed the local charge density to the Gauss-Maxwell expectations \eqref{eq:Maxwell3}.

We observe a difference between positive and negative charges because they move in opposite direction with respect to the electric and magnetic field direction. Radiation feedback is taken into account by a friction term in the Lorentz force, opposite to the velocity, which is proportional to the radiative power $\mathcal{P}$, see eq.\eqref{eq:balance_lorentz_radiation}. We do not use any electron-ion plasma in thermal equilibrium, rather an electron-positron plasma, although the difference between ions and positrons is anecdotal because the particle mass does not intervene in the ultra-relativistic regime, like photons. We also do not have to worry about the fluid motion because particle fill the whole space with a charge density given by Gauss law. This procedure is very similar to its force-free avatar.

In the radiative regime, dissipation of electromagnetic field is controlled by the electric field strength~$E_0$ as measured in the frame where $\vec{E}$ and $\vec{B}$ are parallel because
\begin{equation}
\label{eq:jscalaireE}
\mathbf{j} \cdot \mathbf{E} = |\rho_e| \, ( 1 + 2 \, \kappa ) \, c \, E_0 \geq 0 .
\end{equation}
Expressed in terms of the radiated power we find
\begin{equation}\label{eq:dissipation}
	\mathbf{j} \cdot \mathbf{E} = n_0 \, ( 1 + 2 \, \kappa ) \, \mathcal{P} .
\end{equation}
where $n_0 = |\rho_e/q|$ represents the particle density number required for the minimalistic model of a totally charge separated plasma.

We do not expect reconnection to play any role in the dissipation of the magnetic energy. The radiative Ohm's law behaves as a resistive term. FFE is therefore broken when switching to the radiative solution. All the losses funnel into the particle velocity component along the magnetic field, making particles moving approximately at the speed of light, therefore copiously radiating energy and momentum. The only requirement is that kinetic energy losses being compensated by the electric field work. The balance equation \eqref{eq:balance_lorentz_radiation} connects the radiative losses to the Lorentz force in a stationary state. In this picture we neglect particle inertia compared to the electromagnetic energy and radiative losses. To make an analogy with FFE, particle not only produce the required charge and current density but now they also produce some emission. Their dissipation rate is completely controlled by the radiative term and no reconnection is observed.

In the force-free limit, $E_0$ vanishes and dissipation disappears.
In the minimalistic view, the current~\eqref{eq:J_rad} is imposed only where necessary that is in regions where the condition $E<c\,B$ is violated whether inside or outside the light-cylinder. We call this model the radiative solution (RAD). We could also allow for less dissipation, for instance outside the light-cylinder, in the spirit of the force-free inside/dissipative outside approach of \cite{kalapotharakos_fermi_2017}. We call it force-free inside/radiative outside (FIRO).

\subsubsection{Force-free inside/radiative outside}

Inside the light-cylinder, the corotating electric field~$E$ remains less than $c\,B$. It is therefore always possible to set force-free conditions in this region. However, outside the light-cylinder, the $E$ field can easily surpass the $B$ field strength. In such cases, we can artificially decrease the $E$ strength as in the force-free model. However, as a less stringent method and more realistically, we let the system evolve by adding dissipation not requiring any condition on the $E$ field as allowed by the radiative current prescribed previously in Eq.~\eqref{eq:J_rad}. Therefore in a last regime, we enforce a force-free inside radiative outside (FIRO) model, allowing force-free conditions inside and radiative dissipation outside the light-cylinder.

\subsubsection{Numerical setup}

We performed several sets of runs with the aforementioned three regimes leading to a priori different magnetosphere models. The neutron star radius is set to $R/\rlight=0.3$ and the outer boundary of the simulation sphere is located at $7\,\rlight$ where the light-cylinder is defined by $\rlight=c/\Omega$. This allows us to clearly compute the base of the striped wind on almost one wavelength. The pair multiplicity is set by the user, we chose $\kappa=\{0,1,2\}$. 

The pair multiplicity~$\kappa$ must always be a positive integer. It quantifies the deviation from a purely charge separated plasma. Indeed, in the minimalistic regime, a fully charge separated plasma requires $\kappa=0$. If some weak pair production occurs within the plasma we chose low values such as $\kappa=1,2,5$ or any small integer. For large multiplicities $\kappa \gg 1$, we will show that the solution tends quickly to the force-free magnetosphere when $\kappa$ augments. With $\kappa=2$ the magnetosphere configurations becomes already indistinguishable from the FFE case. The multiplicity $\kappa$ is intimately related to the pair production efficiency within the magnetosphere. PIC simulations have shown that the pair injection process, rate and location, crucially determines the outcome, tending either to a charge separated plasma forming an electrosphere or to an almost neutral plasma leading to a force-free and completely filled magnetosphere. The pair multiplicity remains so far largely unconstrained by observations. However, detailed numerical simulations of pair cascades around the polar caps performed by \cite{timokhin_time-dependent_2010, timokhin_current_2013} showed that values up to $\kappa=10^5$ can be expected. Such high multiplicity leads to an almost perfect force-free regime. Nevertheless, the current prescription containing $\kappa$ as the only free parameter in our model could be supplemented by the freedom in the background charge density $\rho_0$, possibly differing from the standard corotating prescription given by $\rho_{\rm e} \neq \rho_0$. This change in the background dynamics dramatically impacts the magnetosphere electrodynamics. Having no way to constrain this density $\rho_0$ we kept minimal assumption by imposing $\rho_0 = \rho_{\rm e}$.

The pulsar obliquity is denoted by the angle~$\rchi$. We implemented absorbing outer boundary conditions, meaning that the solution becomes unrealistic at distances $r>5\,\rlight$. In the following sections, we derive important quantities related to the pulsar electrodynamics such as its electromagnetic field structure, its spin-down losses, the work done on the plasma and the observational outcome relying on the polar cap shape, their light-curves and the slot gap/striped wind emission properties. A numerical grid of $N_r\times N_\theta \times N_\varphi = 257\times32\times64$ was sufficient for obtaining accurate solutions in all cases.

\begin{figure}
	\centering
	\includegraphics[width=0.95\linewidth]{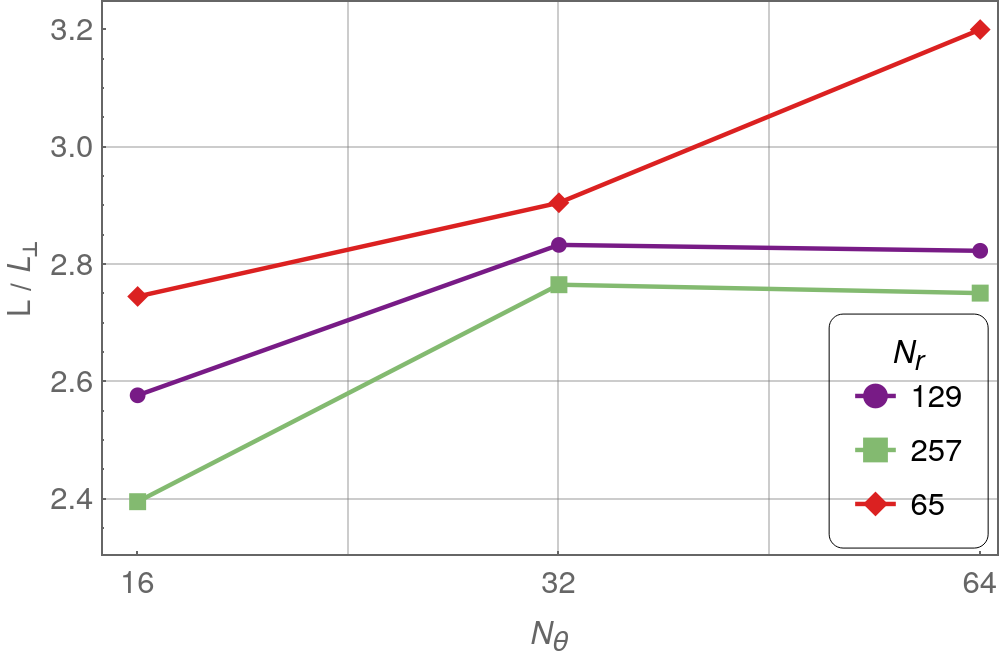}
	\caption{Convergence study for the spin down luminosity~$L$ normalized to the vacuum luminosity~$L_\perp$ of an orthogonal rotator depending on the grid resolution with $N_r$ radial points and $N_\theta$ latitudinal points.}
	\label{fig:convergence}
\end{figure}

The adopted resolution stems from a convergence study of the spin down luminosity. Acceptable accuracy is reached whenever the luminosity at the light-cylinder has converged within 1\%. 
We ran simulations for an orthogonal rotator with several radial and latitudinal grid points from the lowest resolution of $65\times16\times32$ to the highest resolution of $257\times128\times256$. The spin-down luminosity is plotted in Fig.\ref{fig:convergence} and shows that a resolution of $257\times32\times64$ or even $129\times32\times64$ is already sufficient for acceptable accuracy. 
This seemingly low resolution is actually due to the very low dissipation of our fully pseudo-spectral method compared to finite volume or finite difference methods. The grid resolution can be much coarser for spectral methods, especially if the solution is continuous.

Eventually, as a check of our algorithm, including filtering, de-aliasing, absorbing boundary layers and resolution, we computed vacuum solutions comparing our results with expectations from the \cite{deutsch_electromagnetic_1955} solution.

\section{Magnetic field lines}
\label{sec:LigneChamp}

\begin{figure*}
	\centering
	\begin{tabular}{cc}
		\includegraphics[width=\columnwidth]{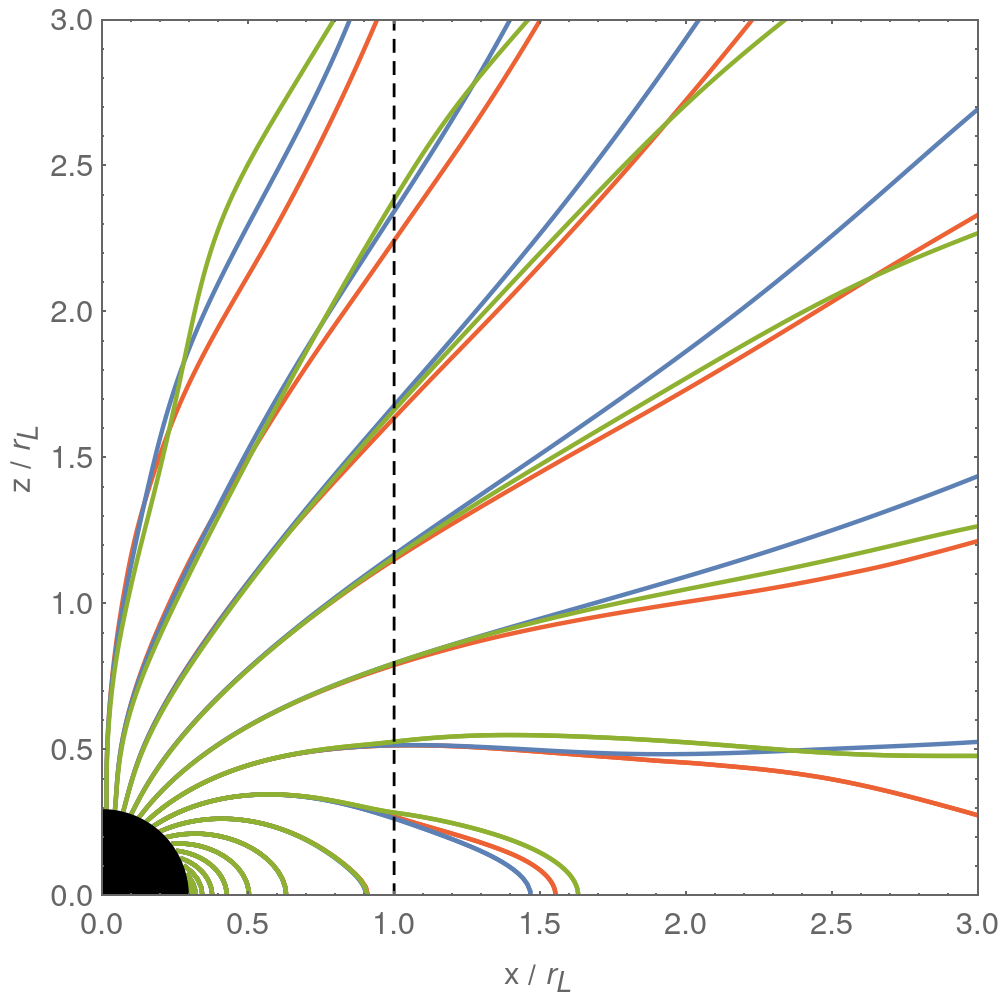} &
		\includegraphics[width=\columnwidth]{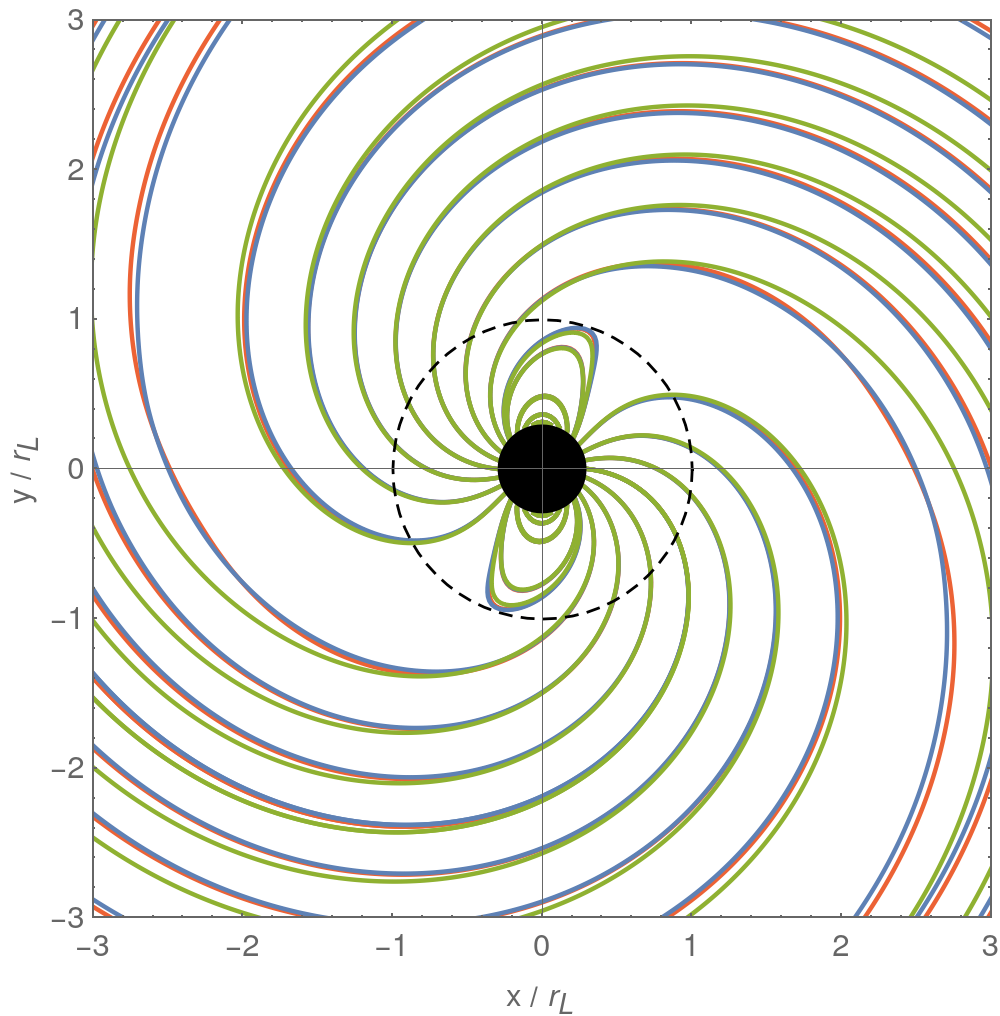} 
	\end{tabular}
	\caption{Magnetic field lines for an aligned rotator, left panel, and for an orthogonal rotator, right panel, in the force-free limit (FFE) in blue, a minimalist radiative magnetosphere (RAD) in red and a FIRO magnetosphere in green, both latter with $\kappa=0$.
		\label{fig:ligne_champ_j}}
\end{figure*}

Electrodynamics of neutron stars relies heavily on its electromagnetic field. We therefore start by showing the magnetic field structure. A full 3D picture being difficult to visualize on a sheet of paper, we restrict ourself to the geometry of magnetic field lines in the meridional plane for an aligned rotator and in the equatorial plane for an orthogonal rotator. Such lines are shown in Fig.~\ref{fig:ligne_champ_j} for the FFE limit and different radiative regimes for an aligned rotator on the left panel and an orthogonal rotator on the right panel. Because the pair multiplicity factor only weakly impacts on the geometry, we only show the cases with $\kappa=0$. Compared to the cases~$\kappa \in \{1,2\}$ we have not found any significant changes, therefore they are not shown in Fig.~\ref{fig:ligne_champ_j}. Inside the light-cylinder, shown as a black dashed line on the left panel and as a circle on the right panel, the magnetic field of the FFE and radiative cases are very similar, whatever the pair multiplicity.
As expected the radiative solutions close more field lines along the equator outside the light-cylinder. 

Next we diagnose quantitatively the effect of a radiative magnetosphere by computing relevant physical quantities such as the spin down luminosity and the work done on the plasma.

\section{Poynting flux}
\label{sec:Luminosite}

A radiative magnetosphere has the interesting property to allow conversion of the Poynting flux into particle acceleration and radiation accounting for the feedback of this current onto the electromagnetic field in a self-consistent manner. In this section, we report the efficiency of Poynting flux decrease depending on the model and on the pair multiplicity. The electromagnetic flux must be compared to the reference situation of a force-free magnetosphere for which no dissipation is expected by construction. However, such solutions develop current sheets that are strongest for an aligned rotator. Numerically such discontinuities are tricky to handle especially for a spectral method where the Gibbs phenomenon easily arises. Some artificial dissipation must be introduced to avoid strong oscillations. However for radiative models, the current prescription naturally leads to some dissipation controlled by a physical parameter like resistivity or radiation damping.

For purely electromagnetic interactions, energy is shared between three quantities, the electromagnetic energy density~$u$ defined by
\begin{equation}
\label{eq:DensiteElectromagnetic}
u = \frac{\varepsilon_0 \, E^2}{2} + \frac{B^2}{2\,\mu_0}
\end{equation}
the Poynting flux defined by
\begin{equation}
\label{eq:FluxPoynting}
 \mathbf{S} = \frac{\mathbf{E} \wedge \mathbf{B}}{\mu_0}
\end{equation}
and the work done on the plasma represented by current density interacting with the electric field at a rate 
\begin{equation}\label{eq:Dissipation_j.E}
\mathcal{D} = \mathbf{j} \cdot \mathbf{E} .
\end{equation}

Particle are assumed to have zero inertia in these simulations. Their velocity is governed by Aristotelian dynamics according to the local electromagnetic field eq.~\eqref{eq:VRR}. In this respect, the lost energy is directly converted into radiation because of this zero lepton mass limit. The strength of the radiative feedback depends on the $E_0$ field which is proportional to the radiated power as shown decades ago by \cite{mestel_axisymmetric_1985}. Magnetic energy is dissipated not via reconnection but via radiation damping, impacting the particle velocity and leading to eq.~\eqref{eq:VRR}.

The dissipative term $\mathcal{D}$ vanishes for a force-free plasma and in the radiation reaction limit it reduces to expression~\eqref{eq:jscalaireE}.
The energy conservation law then reads
\begin{equation}
\label{eq:ConservationEnergie}
\frac{\partial u}{\partial t} + \divg \mathbf{S} + \mathbf{j} \cdot \mathbf{E} = 0 .
\end{equation}
In a stationary state, the electromagnetic energy density~$u$ remains unchanged. Without dissipation, the Poynting flux across a closed surface is conserved but with for instance radiative losses energy flows into the plasma. From the conservation law Eq.~\eqref{eq:ConservationEnergie} integrated within a sphere~$\Sigma$ of radius~$r$ we get
\begin{equation}
\label{eq:Travail}
 \iint_\Sigma \mathbf{S} \cdot \mathbf{e}_{\rm r} \, d\Sigma = - \iiint_V \mathbf{j} \cdot \mathbf{E} \, dV
\end{equation}
where $\mathbf{S} \cdot \mathbf{e}_{\rm r} = S_{\rm r}$ is the radial component of the Poynting flux, $d\Sigma$ a surface element on the sphere and $dV$ a volume element inside the sphere~$\Sigma$.

\begin{figure}
	\centering
	\includegraphics[width=\linewidth]{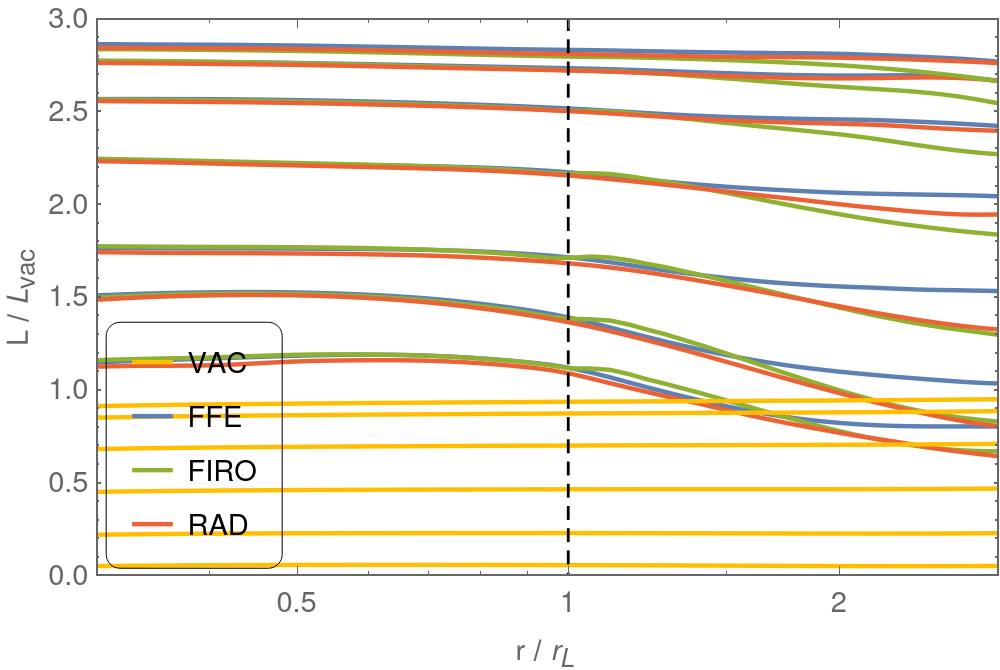}	
	\caption{Radial decrease of the Poynting flux depending on the model. FFE is shown in blue, RAD in red and FIRO in green, both latter for $\kappa=0$. For reference, the vacuum case is also shown in yellow.}
	\label{fig:Luminosite}
\end{figure}

The radial evolution of the Poynting flux is shown in Fig.~\ref{fig:Luminosite} for several models, the force-free (FFE), the force-free inside/radiative outside prescription (FIRO) and the minimalistic radiative approach (RAD). As a check, the vacuum solution (VAC) is also shown to estimate the numerical dissipation. The luminosity is normalized with respect to the vacuum point dipole orthogonal rotator
\begin{equation}\label{eq:spindown_dipole_vide}
L_{\rm vac} = \frac{8\,\upi}{3\,\mu_0\,c^3} \, \Omega^4 \, B^2 \, R^6
\end{equation}
such that the flux plotted is $\ell = L/L_{\rm vac}$.

\begin{figure}
	\centering
	\includegraphics[width=\linewidth]{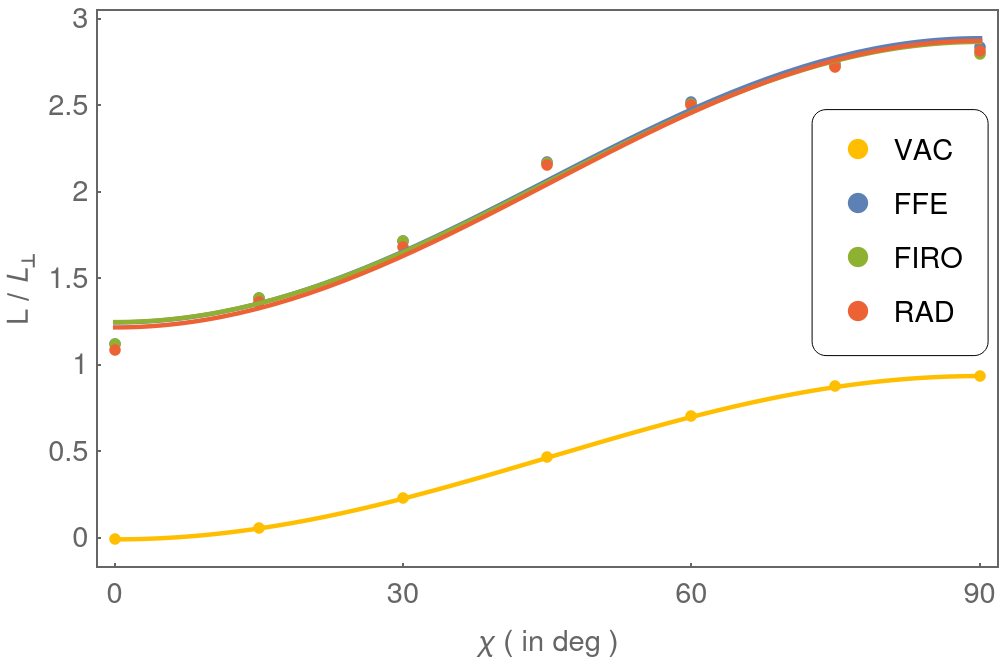}
	\caption{The Poynting flux crossing the light-cylinder for oblique rotators in vacuum (VAC) in yellow, force-free (FFE) in blue, RAD in red and FIRO in green. The solid lines shows the best fit.\label{fig:luminosite_regime}}
	\label{fig:Spindown}
\end{figure}

The angular dependence on the obliquity~$\rchi$ is the same for all regimes, see Fig.~\ref{fig:Spindown}. All the plasma filled fits are well approximated by
\begin{equation}\label{eq:Lapprox}
L / L_{\rm vac} \approx 1.3 + 1.5 \, \sin^2\rchi .
\end{equation}
As a check, for the vacuum case we get
\begin{equation}\label{eq:LapproxVide}
L / L_{\rm vac} \approx 0.96 \, \sin^2\rchi .
\end{equation}
We observe an important dissipation of the Poynting flux for the FIRO case. In order to better localize this dissipative effect within the magnetosphere, we show in the next section the work done on the plasma for an aligned and an orthogonal case.

The sensitivity to the pair multiplicity~$\kappa$ is only weakly perceptible because the prescription for particle injection according to the local electric field via Maxwell-Gauss law already tends to the FFE limit for low multiplicities. What effectively controls the spin down losses and the magnetosphere solution, either closer to an electrosphere or to the FFE regime is the particle density number. PIC simulations have also shown that the injection procedure is critical for the final outcome of the simulation.

\begin{figure}
	\centering
	\includegraphics[width=\linewidth]{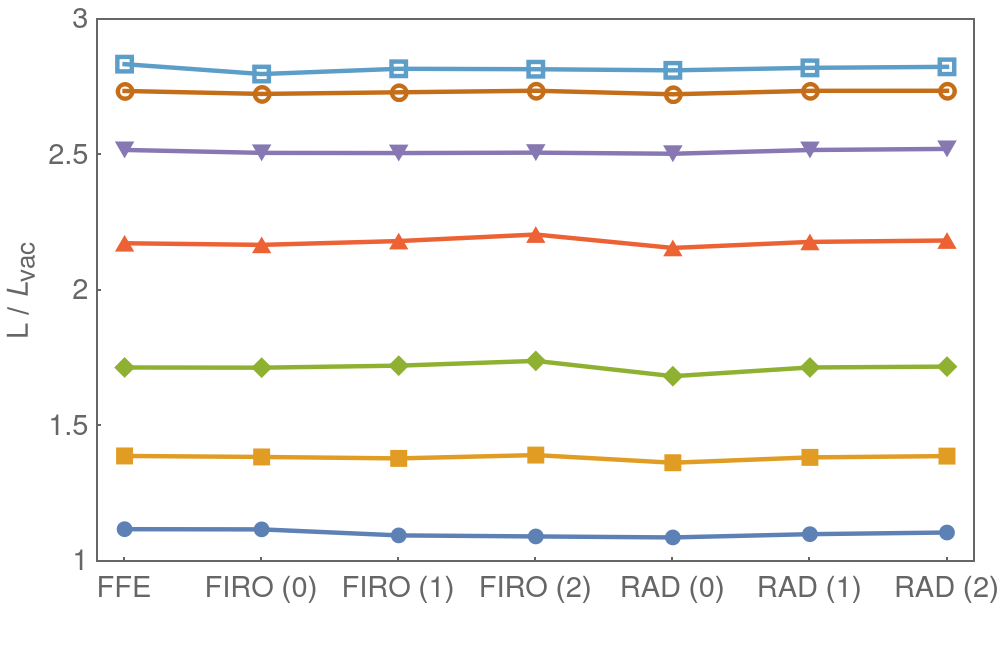}
	\caption{Comparison of spin down luminosities at the light cylinder depending on the plasma model. The integer within the bracket denotes the pair multiplicity factor~$(\kappa)$.}
	\label{fig:comparaison}
\end{figure}

Fig.~\ref{fig:comparaison} shows the different spin down luminosities at the light cylinder depending on the plasma model. The variation in luminosities remains very small except although the radiative solution seems slight more dissipative. The sensitivity to the pair multiplicity is also only weakly perceptible. Most of the dissipation occurs outside the light-cylinder as shown in the next section.

The spin down luminosity is relatively insensitive to the dissipation term because the particle injection scheme follows the force-free scheme by dropping particles where the Maxwell-Gauss law imposes it. Therefore the radiative model tends quickly to a nearly FFE state, even with a low to moderate multiplicity factor~$\kappa$.

\section{Dissipation}
\label{sec:Dissipation}

\begin{figure*}
	\centering
	\begin{tabular}{cc}
	\includegraphics[width=\columnwidth]{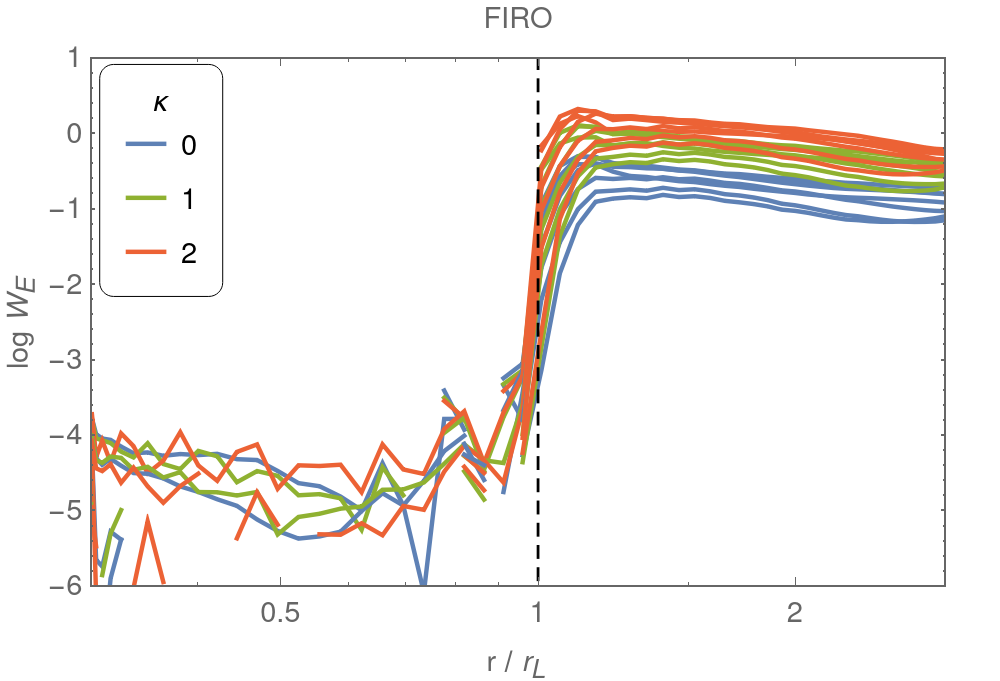} &
	\includegraphics[width=\columnwidth]{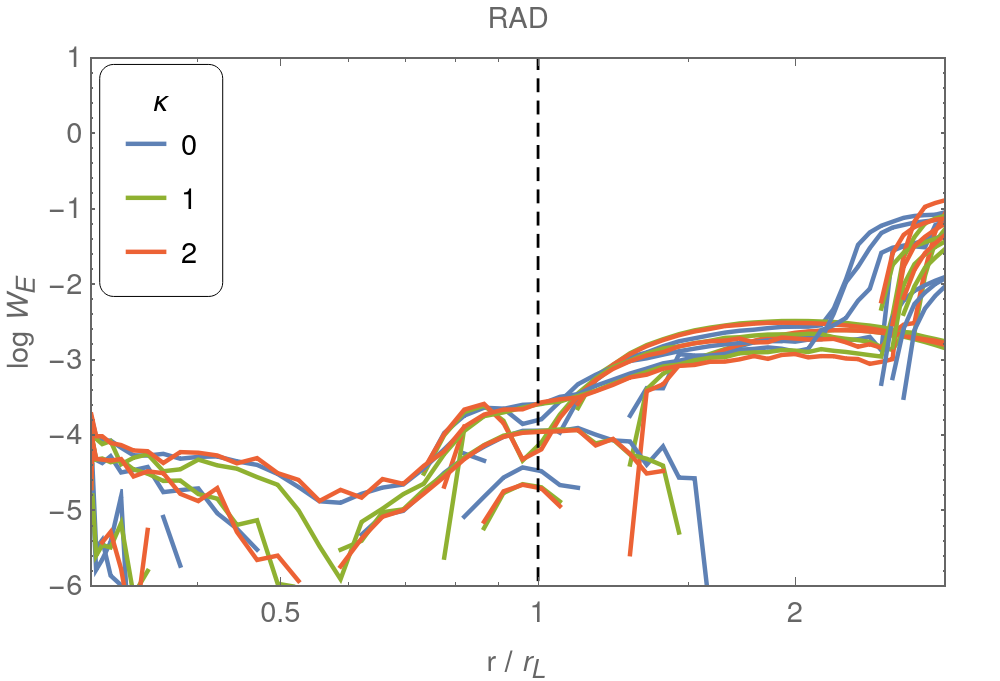}
	\end{tabular}
	\caption{Efficiency of the dissipation according to the integral in Eq.~\eqref{eq:DeriveeTravail} for the FIRO model on the left panel and for the RAD model on the right panel. The pair multiplicity is shown in the legend with different colours. For each multiplicity~$\kappa$, the dissipation is shown for all obliquities in the same colour.}
	\label{fig:dLsdr}
\end{figure*}

Conservation of the total energy implies that some electromagnetic energy went into particle acceleration and radiation. The radial decrease in the Poynting flux~$L$ indicates a sink of electromagnetic energy imputed to the presence of a non ideal plasma. The location where this conversion arises is important for the prediction of observational signatures such as radio and gamma-ray light-curves and spectra. Within a spherical shell of radius~$r$, this dissipation is given by the opposite of the Poynting flux radial derivative as
\begin{equation}
\label{eq:DeriveeTravail}
W_E = \iint_\Sigma \mathbf{j} \cdot \mathbf{E} \, d\Sigma = - \frac{dL}{dr} .
\end{equation}
Fig.~\ref{fig:dLsdr} shows how fast dissipation occurs outside the light-cylinder depending on the radius.
For FFE magnetospheres, dissipation vanishes but because of our numerical filtering procedure and of grid size effects, a small residual work is observed. In the FIRO model, dissipation starts at the light cylinder, increasing to a maximum in the interval $[1,2]\,\rlight$ and then decreases slowly. In the RAD regime, the dissipation sets in with a delay, increasing significantly only beyond a radius $r \gtrsim 2\,\rlight$. This delayed dissipation impacts on the radio time lag of gamma-rays photons with respect to radio photons. We expect to observe a decrease in this time lag for radio loud gamma-ray pulsar, helping to better jointly fit radio and gamma-ray light-curves, as done in \cite{petri_young_2021}.

\begin{figure*}
	\centering
	\begin{tabular}{cc}
	\includegraphics[width=\columnwidth]{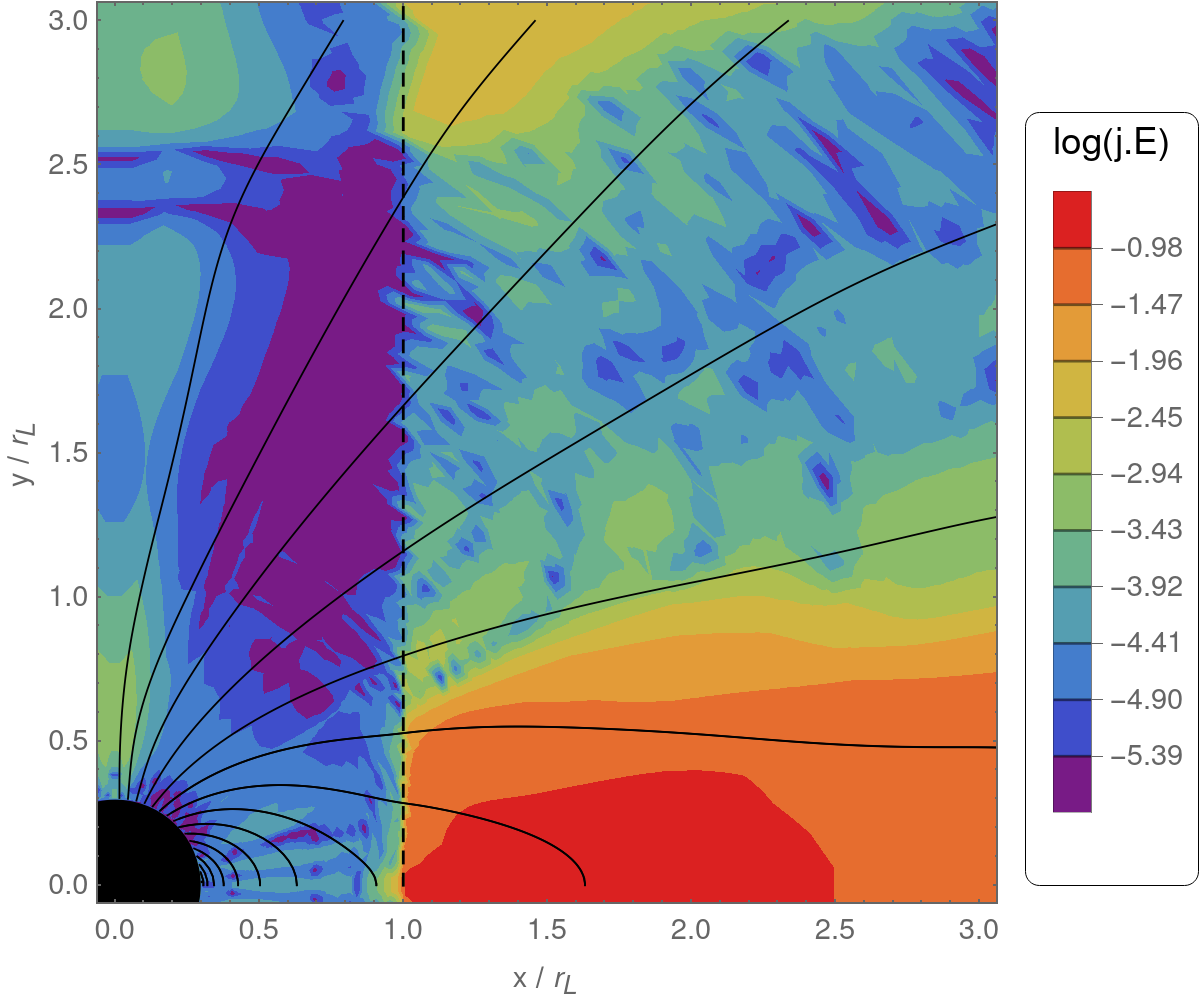} &
	\includegraphics[width=\columnwidth]{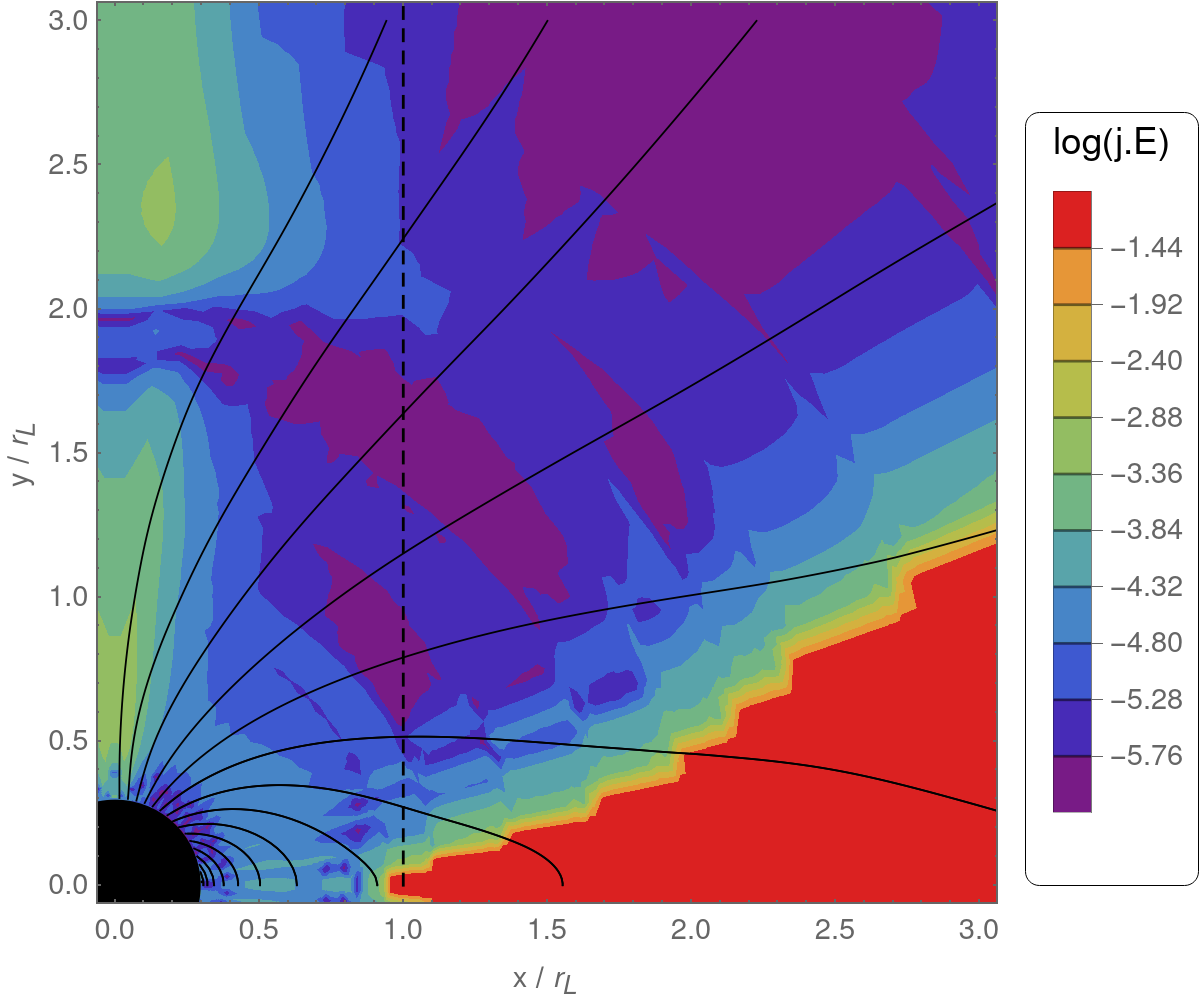}
	\end{tabular}
	\caption{Work done on the plasma for $\kappa=0$ as given by Eq.~(\ref{eq:jscalaireE}), for an aligned rotator with the FIRO model on the left panel and the RAD model on the right panel.}
	\label{fig:dissipation_aligne}
\end{figure*}

\begin{figure*}
	\centering
	\begin{tabular}{cc}
	\includegraphics[width=\columnwidth]{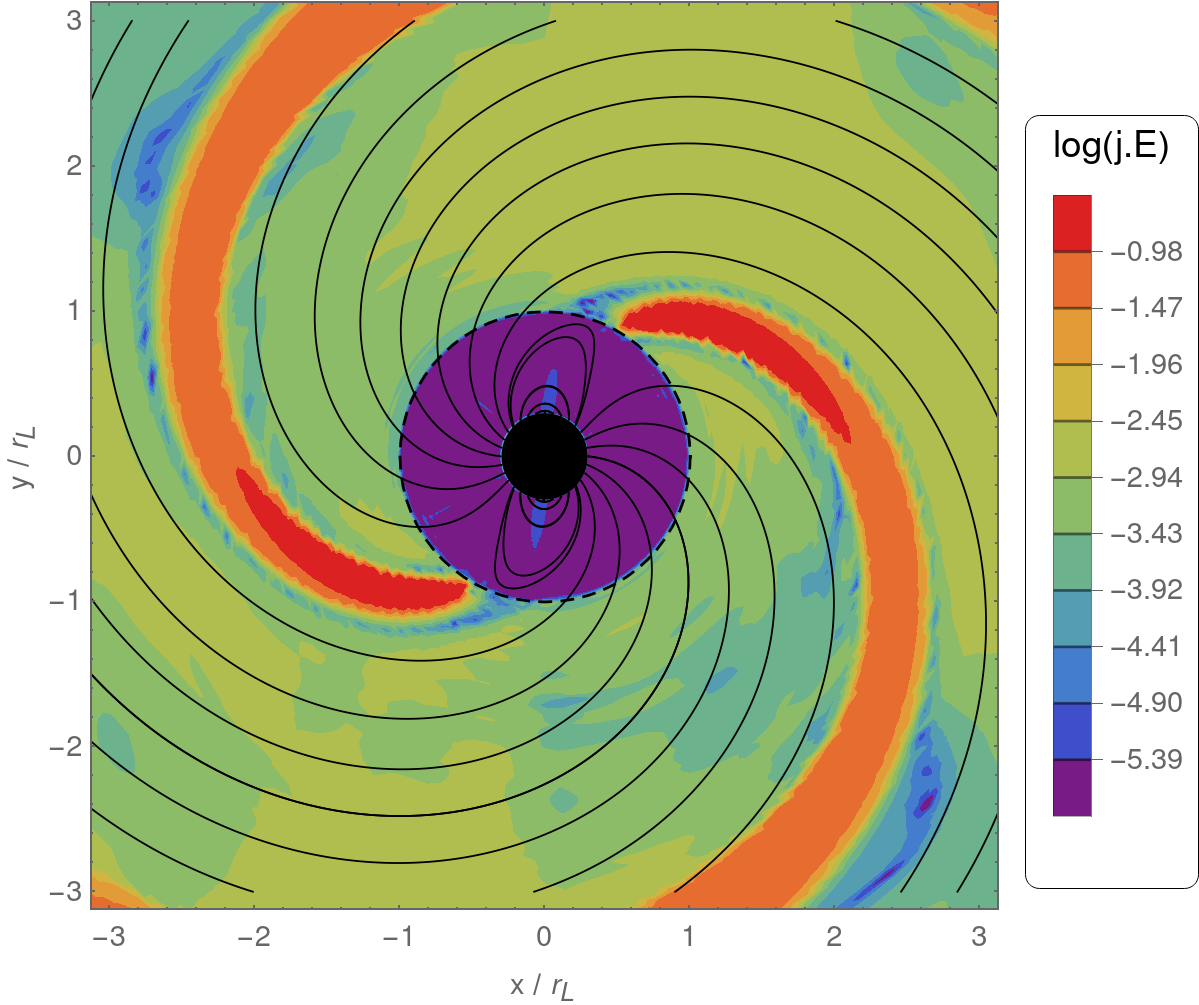} &
	\includegraphics[width=\columnwidth]{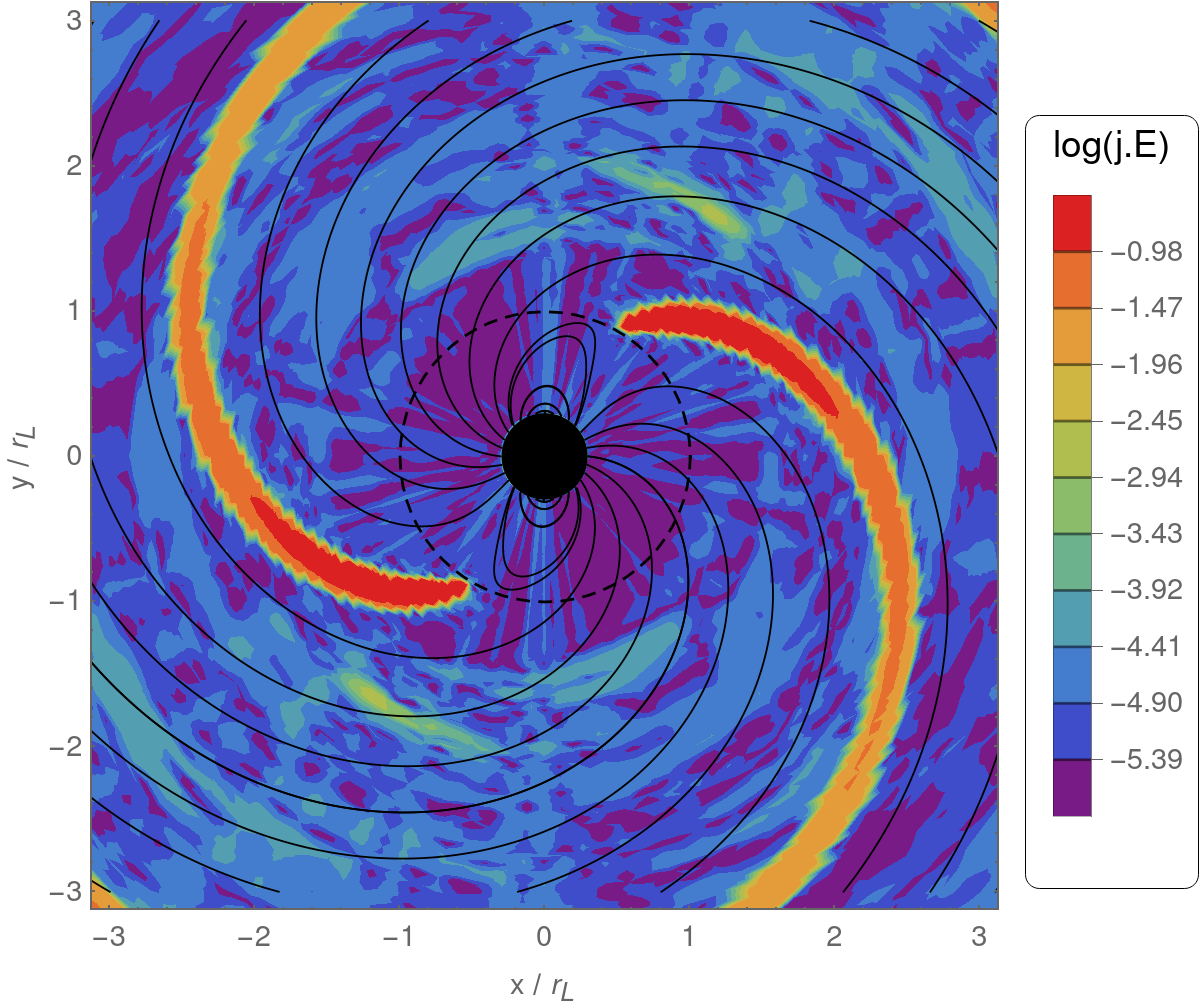}
	\end{tabular}
	\caption{Work done on the plasma for $\kappa=0$ as given by Eq.~(\ref{eq:jscalaireE}), for an orthogonal rotator with the FIRO model on the left panel and the RAD model on the right panel.
	}
	\label{fig:dissipation_orthogonal}
\end{figure*}

In order to better localize the radiative regions where the particle dynamics is the most important, we show maps of the work done locally on the plasma by computing the power defined in Eq.~\eqref{eq:jscalaireE} for aligned and orthogonal rotators. Enlightening cases are shown for FIRO and RAD models in Fig.~\ref{fig:dissipation_aligne} for an aligned rotator and in fig.~\ref{fig:dissipation_orthogonal} for an orthogonal rotator. The Poynting flux flows into the plasma outside the light-cylinder in the vicinity of the current sheet of the striped wind. With increasing distance, the power sharply decreases by two orders of magnitude at the outer boundary because of the decreasing electric field and current density. The thickness of this dissipative region is about $0.2\,\rlight$.

This dissipation layers are the privileged places where high-energy radiation is produced and detected as pulsed gamma-ray emission. This fact is supported by the investigation of radio loud young gamma-ray pulsars for which the rotating vector model is consistent with gamma-ray light-curve fitted for more than a dozen of pulsars \citep{petri_young_2021}.

The acceleration and radiation processes are implicitly implemented by the Aristotelian velocity dynamics. Particles move at the speed of light because of zero inertia approximation and the radiated power is controlled by the $E_0$ field. Particles in the simulations are present but they only contribute to the charge and current density as in the FFE approximation.

In the closed field zone, within the light-cylinder, for the FIRO model, we found a dissipation rate $\mathcal{D}$ which is less than $10^{-4}$ or even $10^{-5}$. Such small values are almost zero from a numerical point of view. The plasma really remains force-free as it should within the numerical error of the algorithm.

The connection between dissipation layers and radiation zones has been confirmed by kinetic simulations such as \cite{philippov_ab-initio_2018} and \cite{chen_filling_2020}.

The current sheet thickness is governed by the local physics, the breakdown of the FFE conditions. It is controlled by the radiative term and not by numerical dissipation which has been checked to remain negligible compared to the dissipation introduced by the radiative Ohm law.

\section{Parallel electric field}
\label{sec:parallel_E}

In order to quantify the presence of a parallel electric field~$E_\parallel$, we plot some maps of the strength of $E_\parallel = \mathbf{E} \cdot \mathbf{B} / B = E_0\,B_0/B$ in the observer frame. This component of the electric field could be responsible for particle acceleration and therefore represents a good indicator of the deviation from force-free conditions. Actually, significant values of $E_\parallel$ are coincident with the dissipation maps shown in the previous section, in Fig.~\ref{fig:dissipation_aligne} and \ref{fig:dissipation_orthogonal}.

\begin{figure*}
	\centering
	\begin{tabular}{cc}
		\includegraphics[width=\columnwidth]{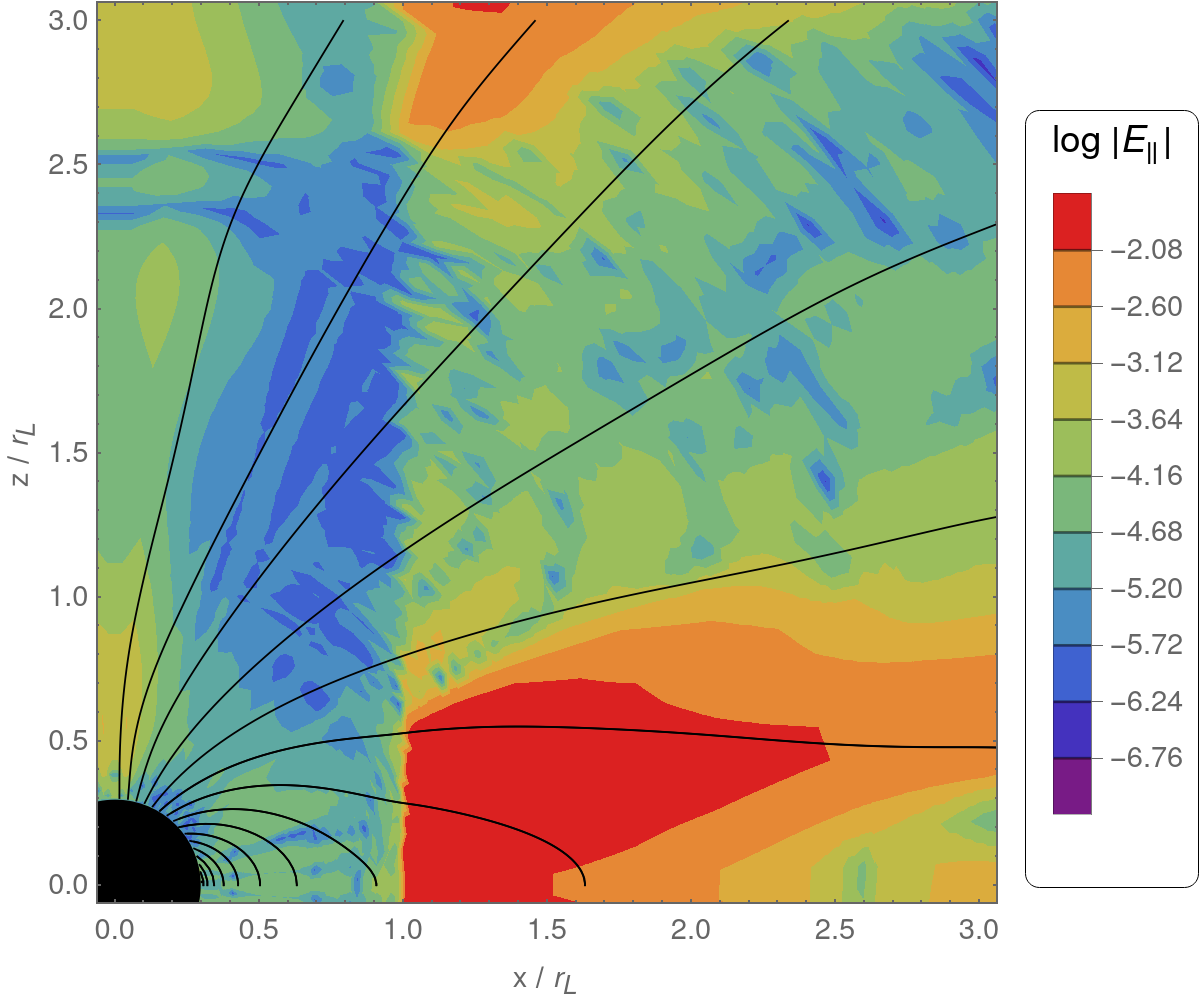} &
		\includegraphics[width=\columnwidth]{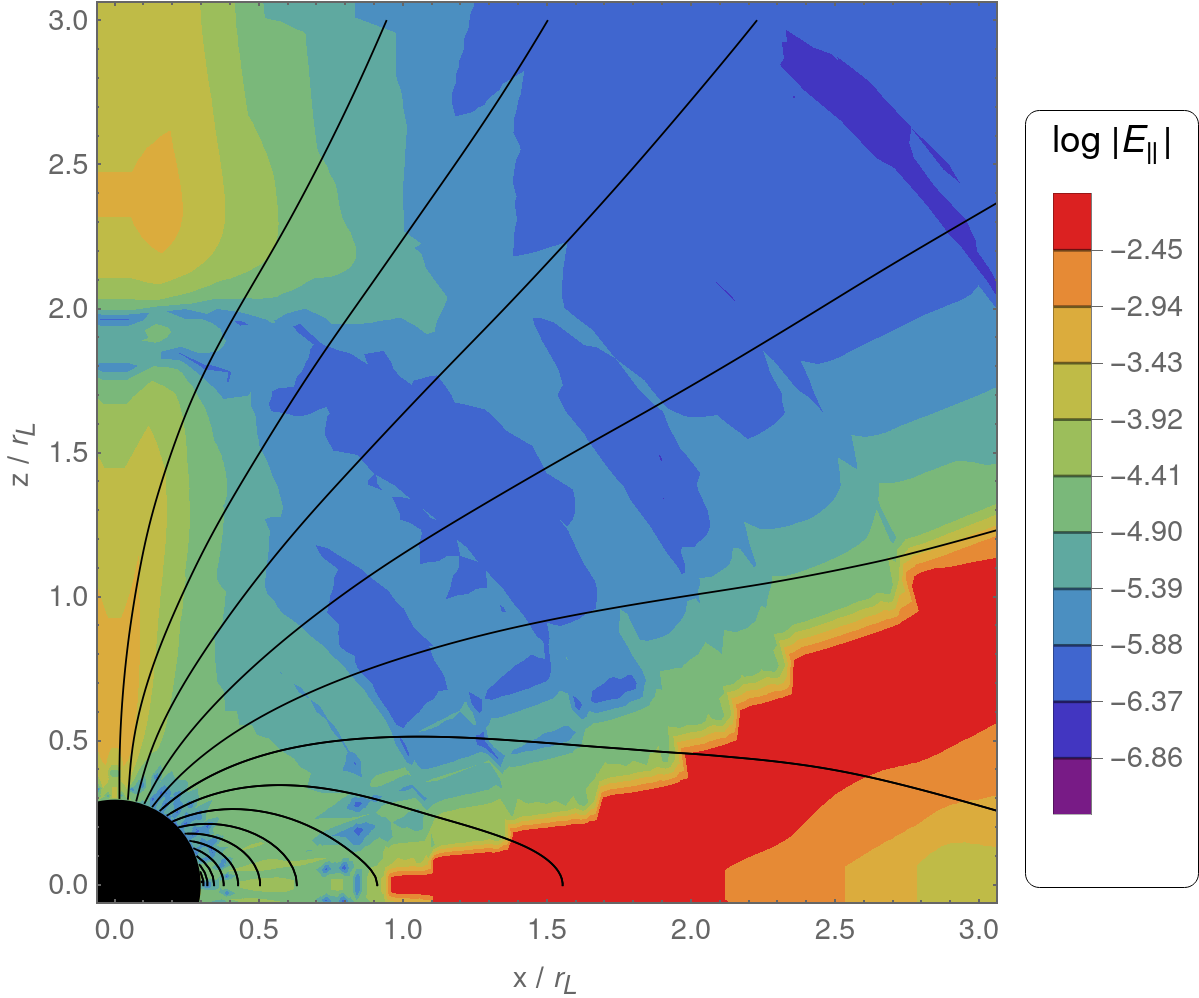} 
	\end{tabular}
	\caption{Absolute value of the parallel electric field~$E_\parallel$ for an aligned rotator, right panel, in the minimalist radiative magnetosphere (RAD) in red and a FIRO magnetosphere in green, both with $\kappa=0$.}
		\label{fig:E_para_aligne}
\end{figure*}

\begin{figure*}
	\centering
	\begin{tabular}{cc}
	\includegraphics[width=\columnwidth]{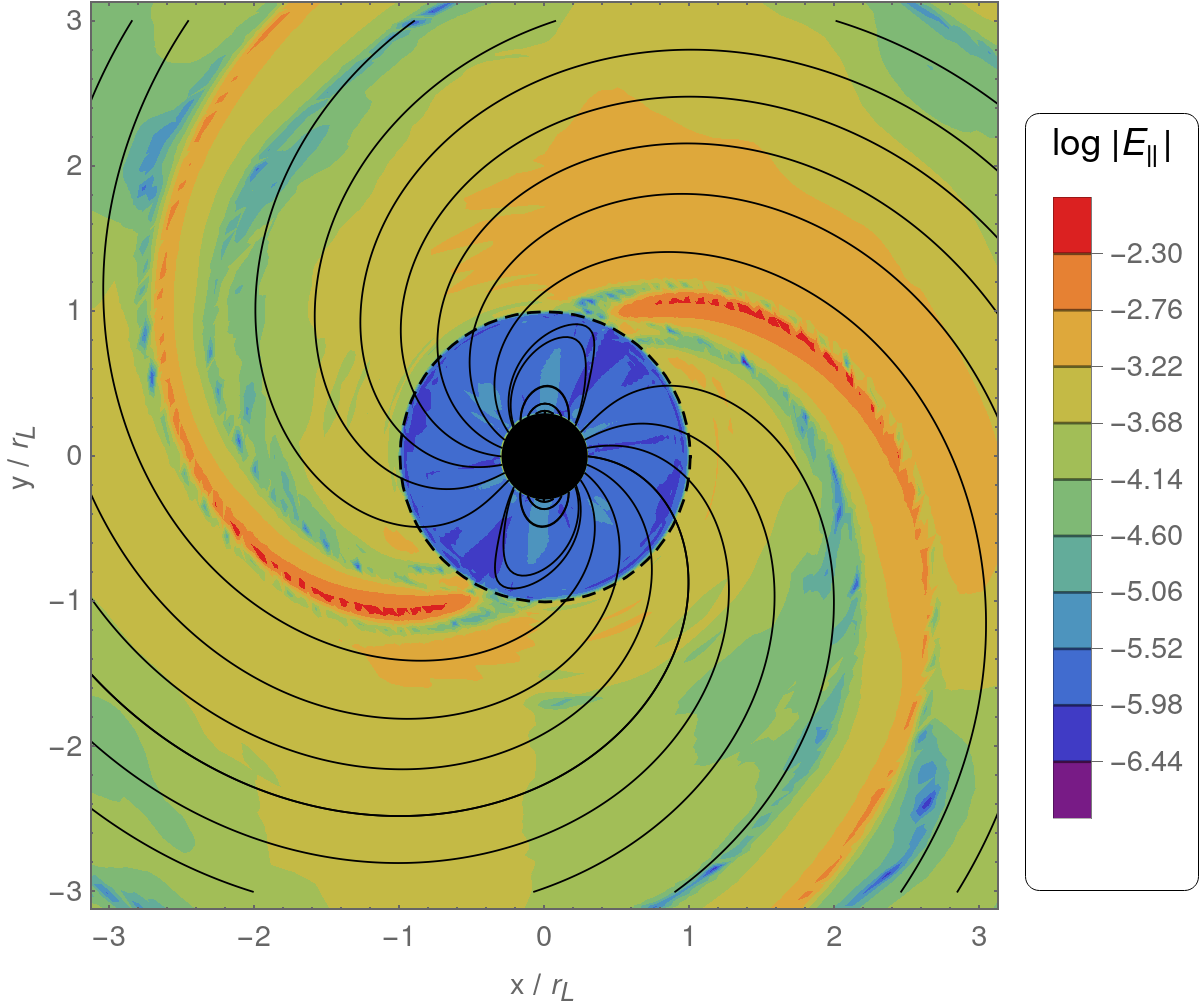} &
	\includegraphics[width=\columnwidth]{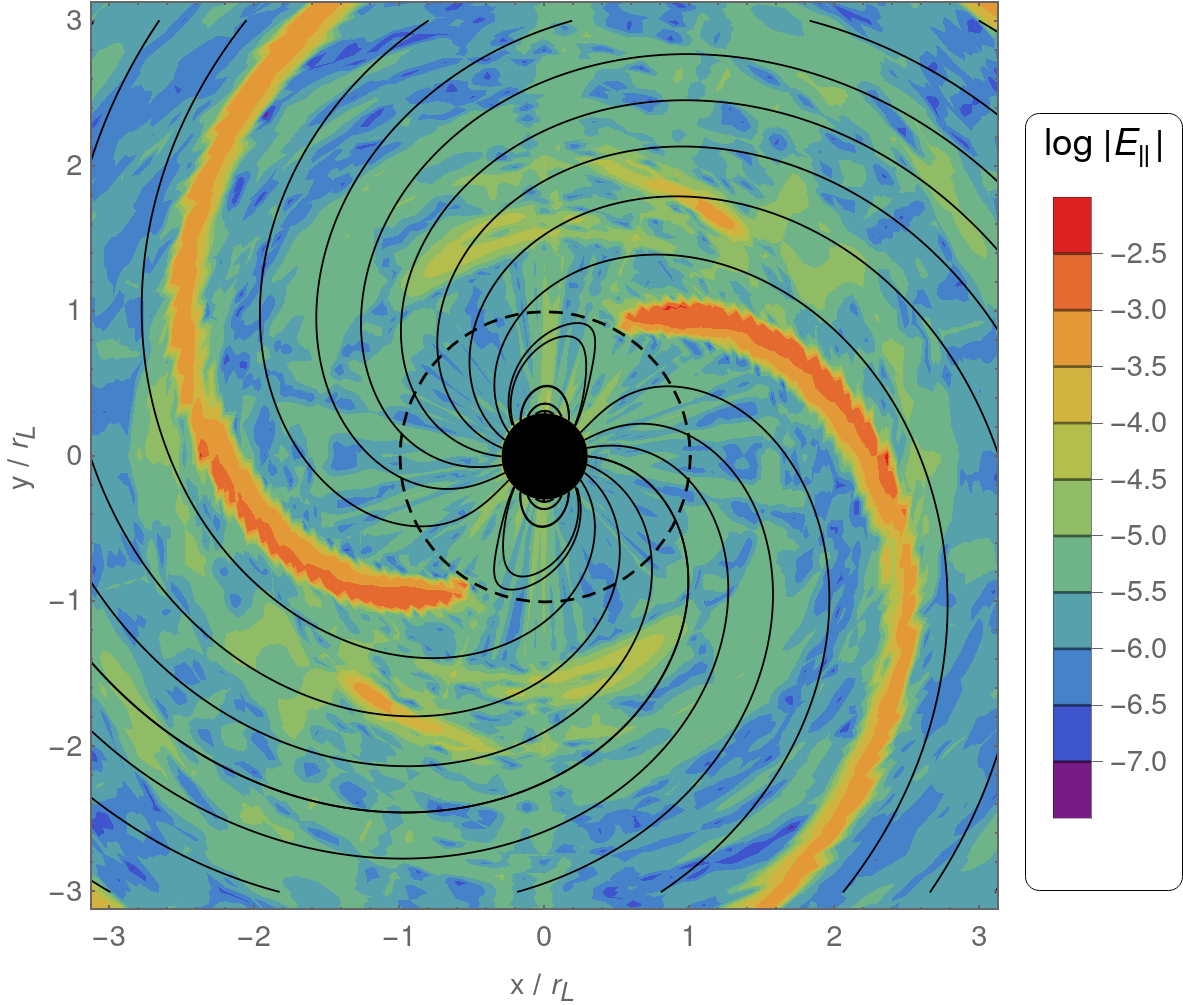}
	\end{tabular}
	\caption{Same as fig.~\ref{fig:E_para_aligne} but for an orthogonal rotator.}
		\label{fig:E_para}
\end{figure*}

Fig.~\ref{fig:E_para_aligne} shows two $E_\parallel$ maps for an aligned rotator with $\kappa=0$, for the FIRO model on the left panel and for the RAD model on the right panel.
Fig.~\ref{fig:E_para} shows the same quantities for an orthogonal rotator.
Particles do not follow field lines any more due to dissipation. They are efficiently accelerated by the electric field~$\mathbf{E}$ along the magnetic field~$\mathbf{B}$ in regions where dissipation is maximal.

Particles are evolved implicitly according to eq.\eqref{eq:VRR}. They follow trajectories imposed by the local electromagnetic field $\mathbf{(E,B)}$ and possess a component along $\mathbf{E}$. This electric field aligned acceleration captures non ideal effects deviating from the pure force-free picture. $\mathbf{E}$ can be larger than $\mathbf{c\, B}$ and there exist a parallel electric field component $E_\parallel$ responsible for these non ideal effects.

There are no strong FFE violation because the particle injection scheme resembles very much to the force-free scheme and therefore the radiative model is still able to tends to a nearly FFE state. The spin down is not much affected but the electric field and the corresponding dissipation rate term eq.\eqref{eq:Dissipation_j.E} are very sensitive to the radiative mechanism however only in very localized areas where $E>c\,B$.

\section{Polar caps}
\label{sec:Calottes}

As a preparation for the investigation of the radio and gamma-ray light-curves of radio loud gamma-ray pulsars detected by Fermi/LAT and related to our radiative magnetosphere, we compute the shape of the polar cap in the different plasma regimes, comparing them to the force-free limit. For better readability with different obliquities $\rchi$, the origin of the plots corresponds to the location of the magnetic north pole, the axes are defined locally for each obliquity by performing a rotation from the rotation axis to the magnetic axis.

\begin{figure*}
	\centering
	\includegraphics[width=\linewidth]{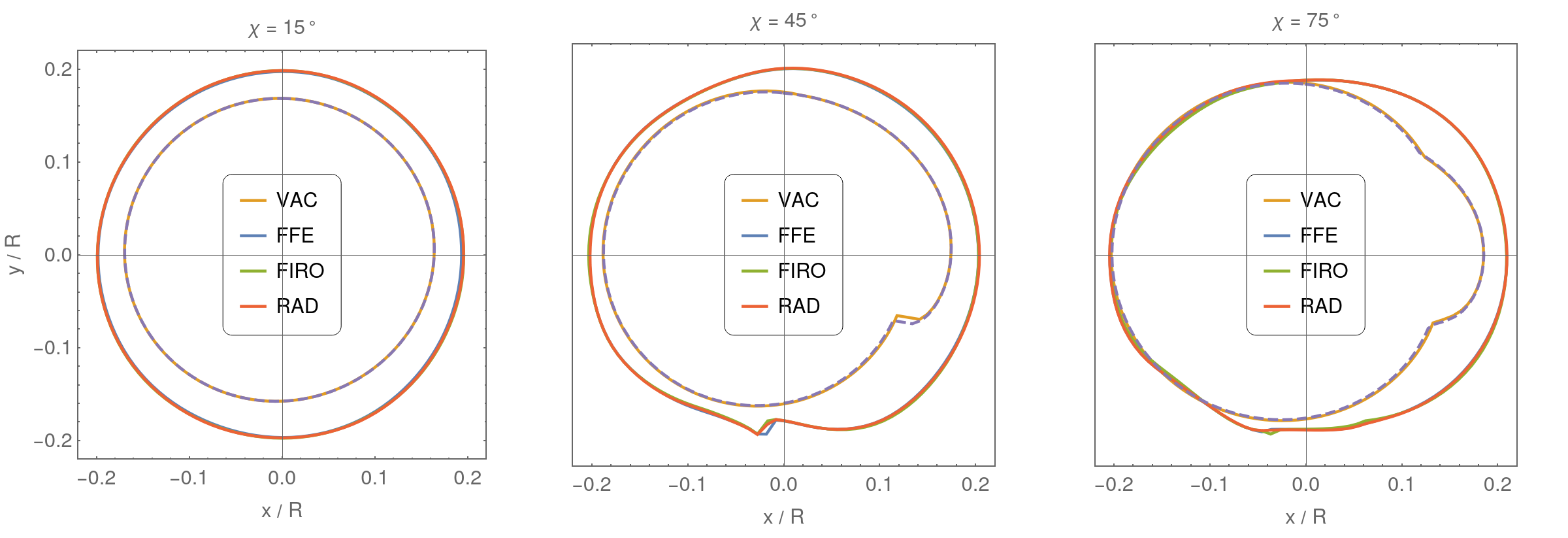} 
	\caption{Polar cap geometry for a rotator with obliquity $\rchi=\{15\degr,45\degr,75\degr\}$ in vacuum (VAC), force-free (FFE) or in the radiation reaction limit, respectively FIRO and RAD. The blue dashed line corresponds to the polar cap computed from the exact Deutsch solution.}
	\label{fig:polar_cap}
\end{figure*}

Illustrative examples are shown in fig.~\ref{fig:polar_cap} for the polar cap rim with $\rchi = \{15\degr,45\degr,75\degr\}$, in vacuum (VAC), force-free (FFE), FIRO and RAD regimes. The vacuum polar cap shapes computed from our pseudo-spectral simulations are shown in orange solid line and checked against those polar caps found from the exact analytical Deutsch solution and shown in dashed blue lines. The agreement between both contours is excellent and gives us confidence about our results for plasma fields magnetospheres. The polar caps for the FFE, FIRO and RAD regimes are also shown in fig.~\ref{fig:polar_cap}, respectively in blue, green and red solid line. We have not noticed any significant change in these caps and their rims almost overlap whatever the regime.
Nevertheless, compared to vacuum, the area of these polar caps is larger than in the vacuum case because magnetic field lines open up due to the magnetospheric current. The presence of the plasma inflates the caps.

We conclude that the impact of radiation on the polar cap shape remains rather weak when compared to the force-free solution. Even if the ideal plasma approximation does not produce any parallel electric fields that could accelerate particles and therefore does not produce any radiation, from a geometrical point of view, it nevertheless offers a reasonably faithful picture of the electromagnetic field of a dissipative magnetosphere, wherever the radiative dissipation occurs, everywhere outside the light-cylinder or only where required by local conditions imposed by the electromagnetic field. We stress that within a fluid model, as the one employed in our study, the acceleration mechanism cannot produce power law distribution functions. It only heats up particles by keeping them thermal. This acceleration is accounted for by the parallel electric fields~$E_\parallel$ component for a resistive plasma for which magnetic energy is dissipated into particle kinetic energy. This requires an Ohm's law like eq.\eqref{eq:J_rad} which allows the presence of a parallel electric field~$E_\parallel$. Injecting test particles into this configuration would offer a good compromise between a fully kinetic simulation and a fluid approximation, leading to possible non thermal acceleration and building power law distribution functions for those particles.

The observation that force-free and dissipative magnetospheres look rather similar relies mainly on the fact that the particle injection scheme derives from Maxwell-Gauss law, sowing particles locally at the rate imposed by the electric field. In such an approach, a fully charge separated plasma present in the whole magnetosphere deviates only slightly from the force-free counterpart. This conclusion is not an artefact from the grid resolution which could impact on the physical dissipative term but a natural consequence of the lepton injection method. Replacing the charge density derived from Maxwell-Gauss by another explicit spatio-temporal injection dependence would lead to more effective dissipation. But this is at the expense of adding more arbitrariness into the model that we wanted to avoid in the present investigation.

Small changes on the stellar surface are amplified at the light-cylinder and beyond, therefore we could expect a significant change in the multi-wavelength light-curve predictions. This last point connecting our simulations to observations is touched in the next section through comparison of sky maps in the radio and the gamma-ray band.

\section{Pulsed emission}
\label{sec:emission}

Dissipative magnetospheres are necessary to produce some radiation as detected by a distant observer. The location and geometry of the emission regions strongly imprint on the multi-wavelength light-curves and the phase-resolved spectra. In this last section, we compute sky maps and light-curves for the aforementioned models, highlighting the differences expected depending on the plasma regime, ideal or radiative.

We consider the three main emission regions to be, first the polar cap for radio photons, second the slot gap for high-energy gamma-ray photons and third the striped wind model for high and very high-energy gamma-rays up to the TeV range. For radio photons, we assume a polar cap model with emissivity shaped by a Gaussian function centred on the magnetic axis. For gamma-ray photons, we assume a striped wind model with emissivity starting at the light-cylinder and focused along the current sheet or a slot gap extending from the surface (meaning here 0.3~$\rlight$) up to the light-cylinder. Some more details about these emission models can be found in \cite{petri_general-relativistic_2018}.

In our model, the size of the radio cone emission is controlled by the polar cap rims computed in the previous section. We assume that photon are produced with an altitude in the interval $[0.3,0.4]\,\rlight$. In this region close to the stellar surface, the electric field remains weak compared to the magnetic field and particles follow almost magnetic field lines. Photons are therefore shoot in a direction tangential to the magnetic field lines, including retardation and aberration effects. Moreover, we implemented a Gaussian profile centred along the magnetic moment axis with a typical width equal to the cone supported by the last open field lines. We adopted this picture in order to mimic the true radio profiles observed in many pulsars.

Gamma-ray photons are produced in the equatorial current within the striped wind, outside the light-cylinder. Because the wind is expanding almost radially, these photons are emitted in the radial direction. The emissivity is maximal at the centre of the current sheet and decreases following another Gaussian shape when deviating from this sheet. The emission zone extend from $1\rlight$ to $3\rlight$. Much more details can be found in \cite{benli_constraining_2021} and \cite{petri_young_2021} and references therein.

In order to probe the radio emission mechanism from the polar cap and the gamma-ray emission from higher altitudes, kinetic physics is required to capture the non-ideal electric field, the gap formation, and the pair production. However such study is out of the present scope. Attempts to better understanding the radio emission generation have been pursued by for instance \cite{philippov_origin_2020} and for the gamma-ray emission by for instance \cite{kalapotharakos_fermi_2017}. Here we are only interested in the radio and high energy pulse profiles implied by the geometrical configuration of radiative magnetospheres. We occult the detailed energetics of individual particles, radiation and electromagnetic interactions.

\subsection{Sky maps}

\begin{figure}
	\centering
	\includegraphics[width=\linewidth]{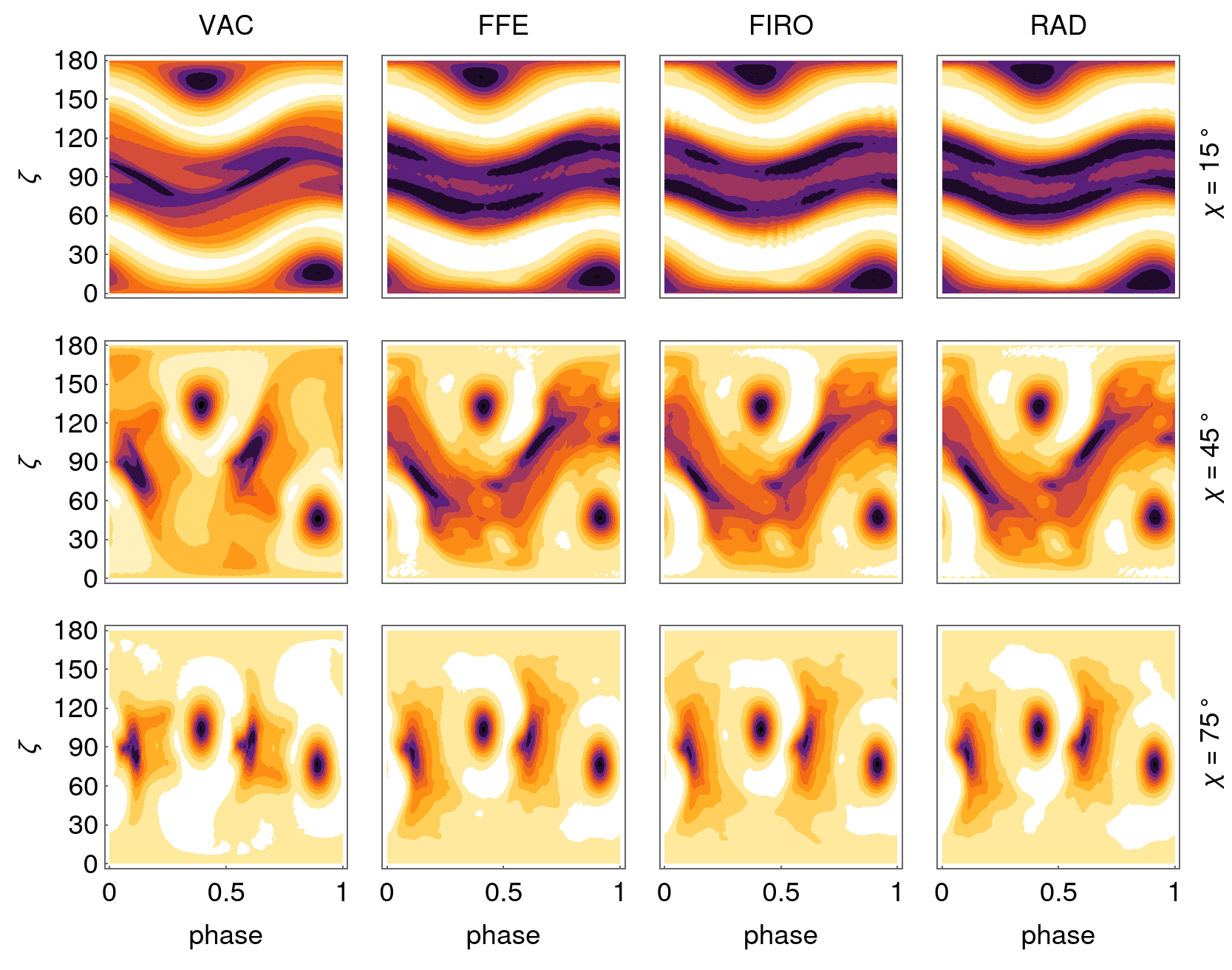}
	\caption{Sky maps of the ideal and radiative models compared to the vacuum case. From left to right column: vacuum (VAC), force-free (FFE), FFE inside/radiative outside (FIRO), and radiative (RAD) on the right. Each line represents a different obliquity given from top to bottom by $\rchi = \{15\degr,45\degr,75\degr\}$. The gamma-ray emission assumes a striped wind model.}
	\label{fig:carte_complete_vent}
\end{figure}

\begin{figure}
	\centering
	\includegraphics[width=\linewidth]{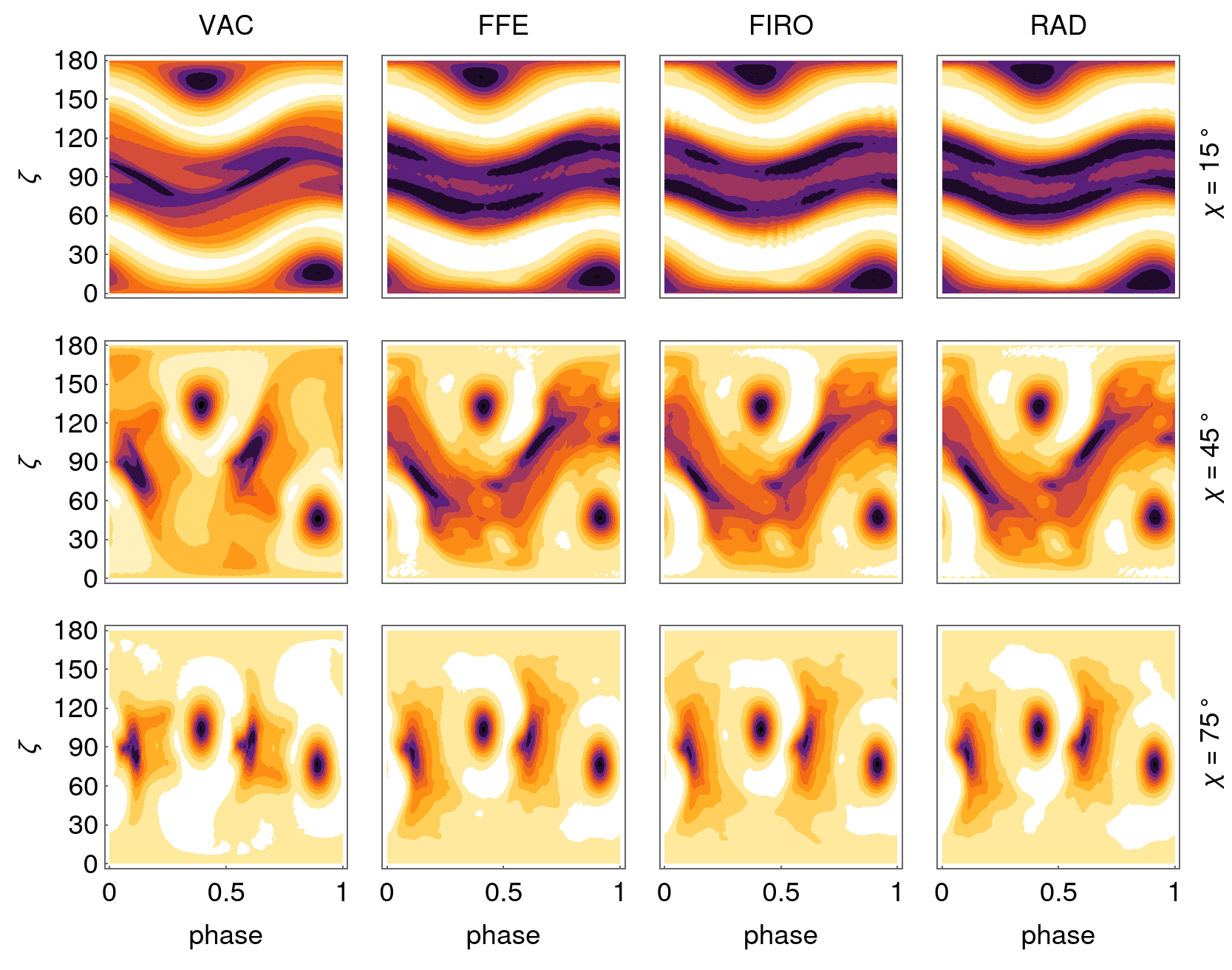}
	\caption{Same as fig.~\ref{fig:carte_complete_vent} but for the gamma-ray emission assumes a slot gap model.}
	\label{fig:carte_complete_cavite}
\end{figure}

We start by reckoning a full set of light-curves in radio and high-energy, following the three emission regions. The combined polar cap/striped sky maps are summarized in Fig.~\ref{fig:carte_complete_vent} for $\rchi = \{15\degr,45\degr,75\degr\}$ and the combined polar cap/slot gap is summarized in Fig.~\ref{fig:carte_complete_cavite} for the same obliquities. It is also instructive to show the expectations from the Deutsch vacuum solution. The observer line of sight inclination~$\zeta$ varies from 0\degr to 180\degr. Each column in this plot depicts a particular model. From left to right column we have successively the vacuum (VAC), the force-free (FFE), the FFE inside/radiative outside (FIRO), and the radiative (RAD) regimes. Each line represents a different obliquity given from top to bottom by $\rchi = \{15\degr,45\degr,75\degr\}$.

Interestingly, the force-free maps are very similar to the radiative maps. 
The FFE case and radiative runs are different in their details. Radiative models clearly show some regions where the electric field exceeds the magnetic field $E>c\,B$ and some dissipative regions where $\mathcal{D}$ is non negligible. Whereas the global magnetosphere is not drastically impacted by the local dissipative terms, the sky maps are.

Even the vacuum solution already produces maps resembling the plasma filled cases, especially when looking close to the equator ($\zeta \approx 90\degr$). Inspecting more carefully Fig.~\ref{fig:carte_complete_vent} and Fig.~\ref{fig:carte_complete_cavite}, we conclude that the actual plasma regime only weakly impacts on the sky maps, whether in radio or in gamma-rays. From a geometrical point of view, when investigating light-curve shapes, the force-free limit allows for an accurate study of pulse profile without adding any free parameter into the game. The geometric dependence on $\rchi$ and $\zeta$ is already faithfully reproduced in FFE. However, when energetic considerations come into play, the radiative models will generate very different phase-resolved spectra and multi-wavelength light curves because of the varying parallel electric field acting on the particle dynamics. This requires deeper investigation of particle acceleration and radiation that we leave for future work.

\subsection{Light curves}

\begin{figure*}
	\centering
	\includegraphics[width=0.95\linewidth]{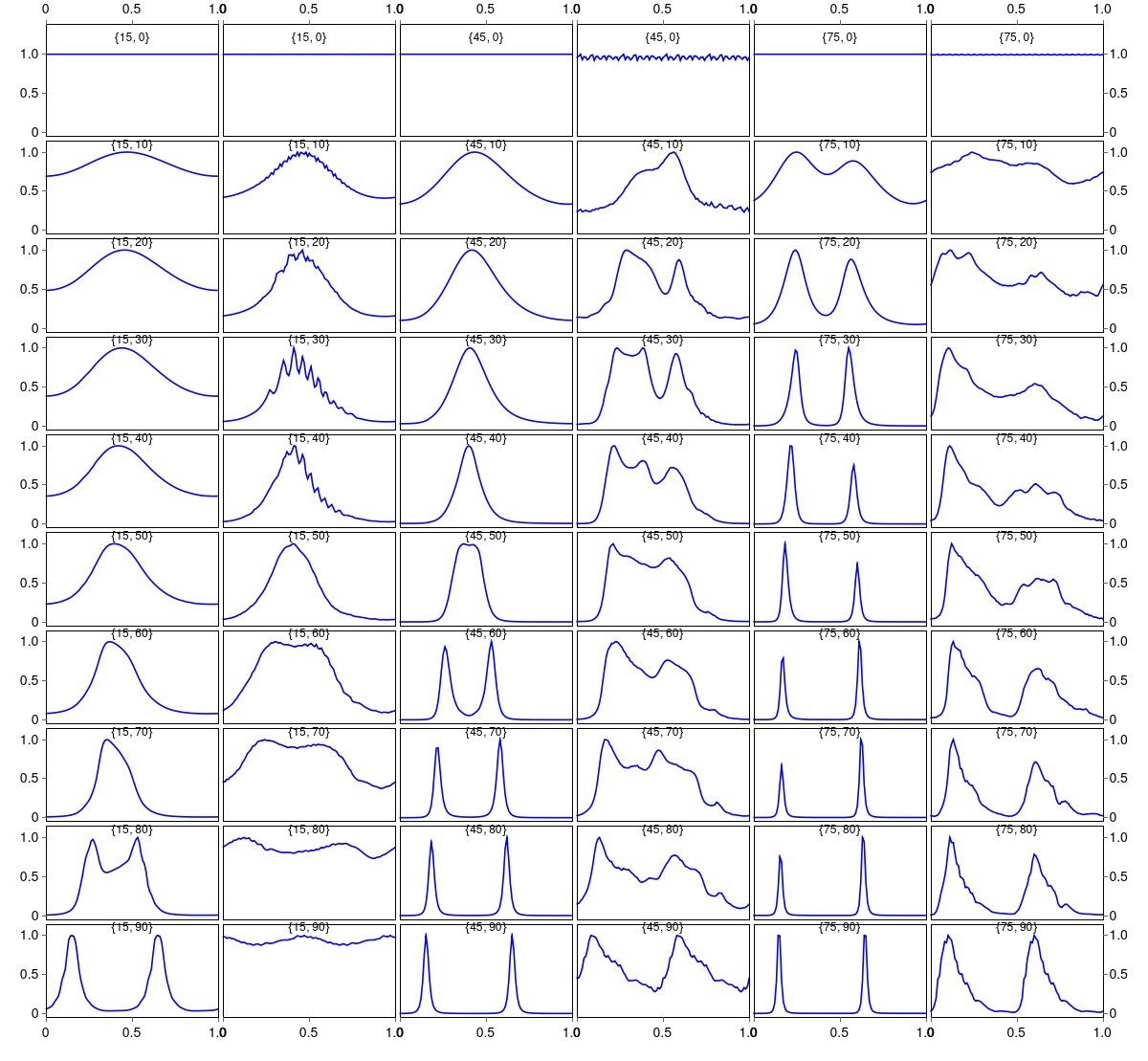}
	\caption{Atlas of striped wind and slot gap light-curves for $\rchi = \{15\degr,45\degr,75\degr\}$ and $\zeta\in[0\degr,90\degr]$ in steps of 10\degr, see the inset in the format $\{\rchi, \zeta\}$. The first, third and fifth column are for the striped wind whereas the second, fourth and sixth column are for the slot gap.}
	\label{fig:atlasgamma}
\end{figure*}

As a typical example of different light curves constructed from these models, we plot an atlas of gamma-ray light curves in Figure~\ref{fig:atlasgamma} for $\rchi=\{15\degr,45\degr,75\degr\}$ and in steps of 10\degr for $\zeta\in[0\degr,90\degr]$, according to the striped wind and the slot gap model in the force-free limit.

\begin{figure}
	\centering
	\begin{tabular}{ccc}
	\includegraphics[width=0.45\columnwidth]{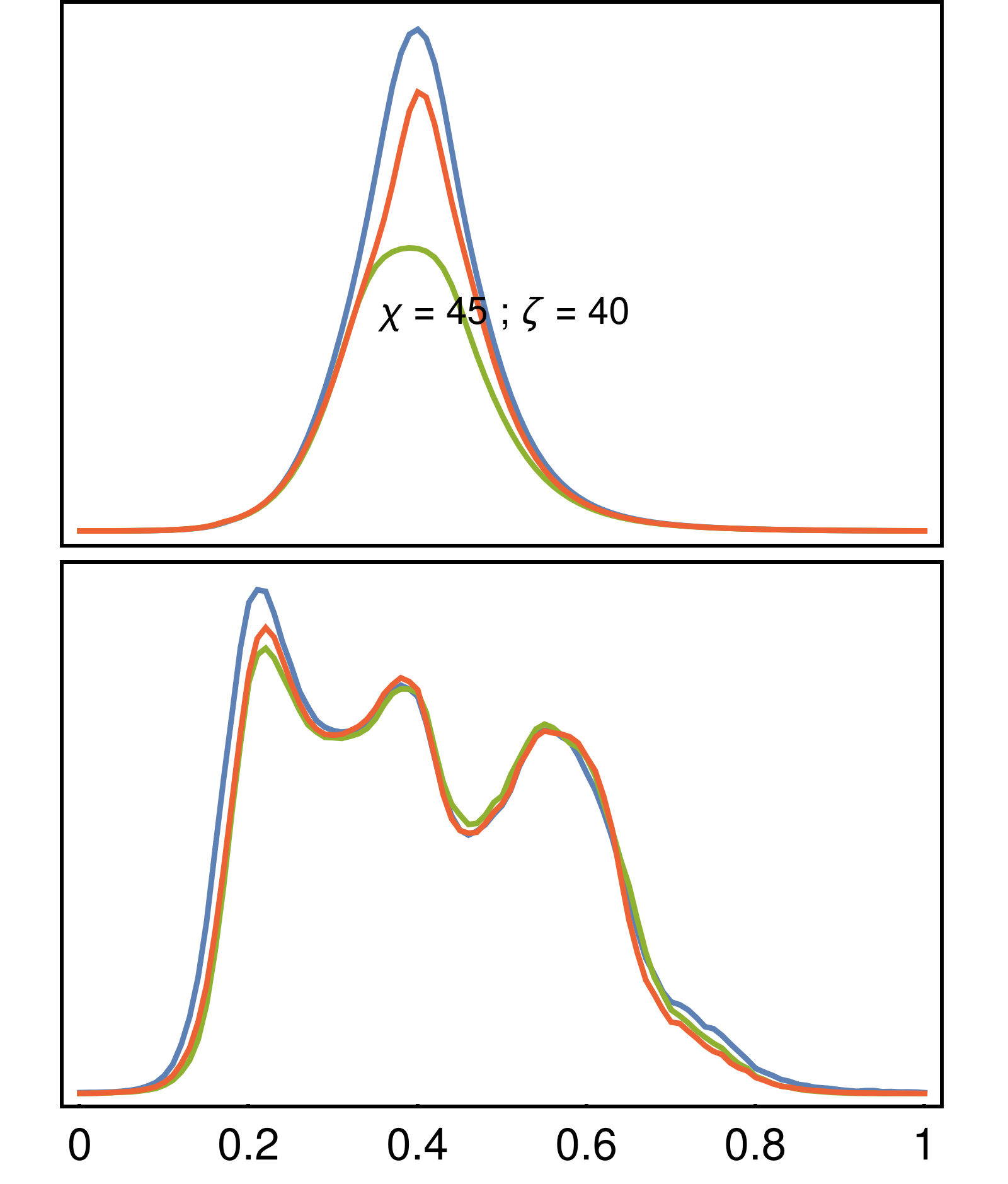} &
	\includegraphics[width=0.45\columnwidth]{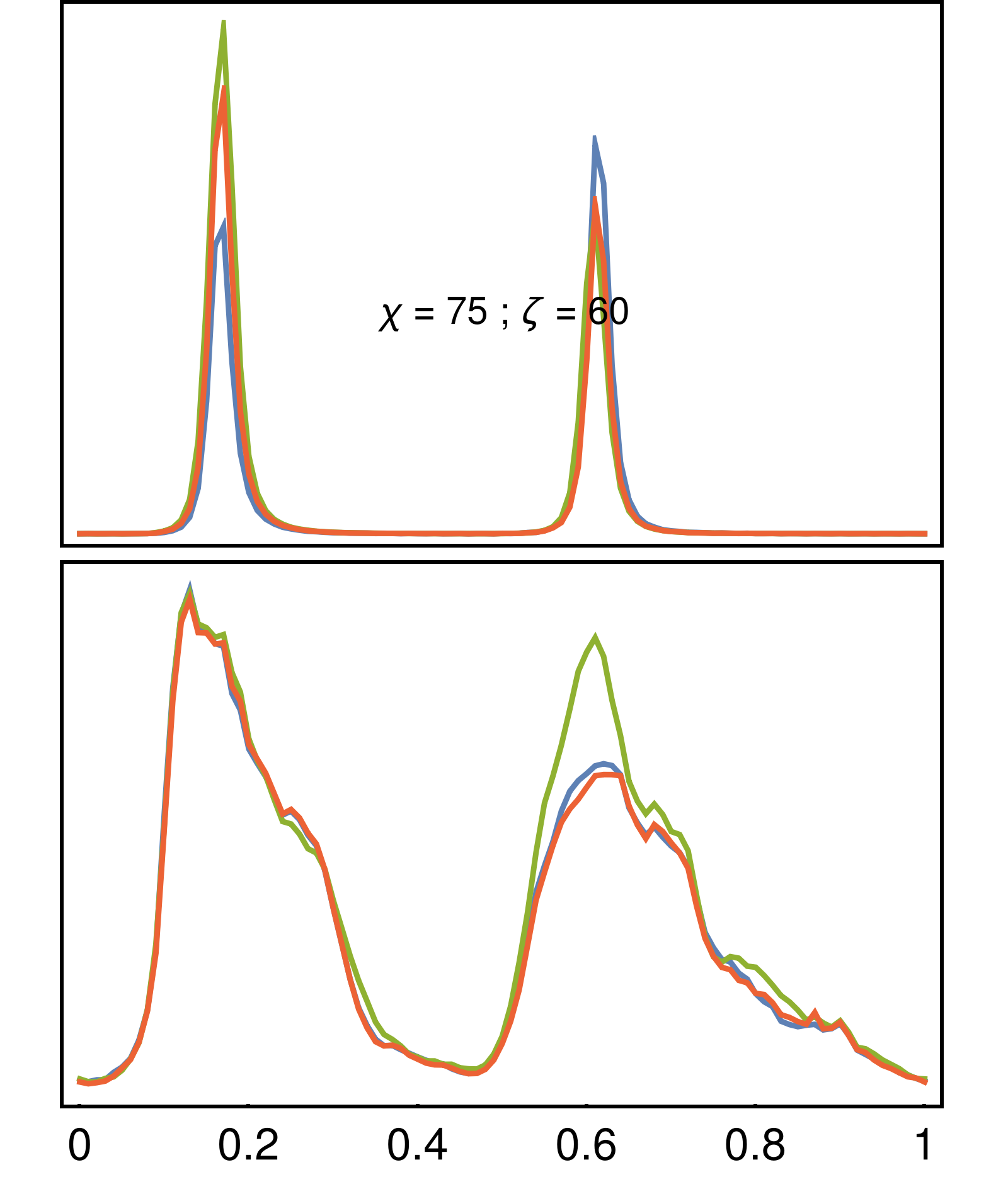} 
	\end{tabular}
	\caption{Examples of gamma-ray light curves extracted from the sky maps shown in figure~\ref{fig:carte_complete_vent} and \ref{fig:carte_complete_cavite} with the geometry given in the inset. On the top panel the results for the striped wind and on the bottom panel for the slot gap.}
	\label{fig:courbe_lumiere}
\end{figure}

For the plasma filled models, we only notice a variation of several percent in phase lag between the light curves. To a large extent, the double peak gamma-ray separation remains insensitive to the model used. Discrepancies in phase are difficult to detect in real gamma-ray observations. However, more importantly are the variations in the peak maximal intensity between the models, notably the reversal of the dominant peak, leading or trailing with respect to radio, when switching from FFE/FIRO to RAD model, see Fig.~\ref{fig:courbe_lumiere} with $\{\rchi,\zeta\}=\{45\degr,40\degr\}$ on the left panel and $\{\rchi,\zeta\}=\{75\degr,60\degr\}$ on the right panel.
The slot gap model produces much wider profiles, broader than what is observed by Fermi/LAT such that we favour the striped wind to explain GeV light-curves although for some pulsar showing a kind of plateau emission the slot gap would better fit the observations. As a conclusion, we think that the majority of the gamma-ray pulsars fall into the striped wind emission model except for some outliers possibly also dominantly emitting in the slot gap sites.

\section{Conclusions}
\label{sec:Conclusion}

Finding self-consistent dissipative pulsar magnetospheres is important to localize possible sites for their broadband pulsed emission. The underlying processes of particle acceleration and radiation remain to be properly identified. In this paper, we constructed self-consistent radiative pulsar magnetospheres in the radiation reaction regime where ultra-relativistic particles flow around the neutron star. We introduced partially and minimalistic radiative models, demonstrating that the global electromagnetic topology remains mainly insensitive to the actual dissipation regime. In any case, most of the radiation occurs in the current sheet outside the light-cylinder but close to it.

Although different models produce marginal differences in the gamma-rays and radio light-curves, we believed that current observations especially in high-energy are not sensitive enough to disentangle for instance the minimalistic from the partially radiative magnetosphere. For better comparison with observations, the computation of phase-resolved spectra and multi-wavelength light-curves will undoubtedly help to segregate between competing dissipative models like the radiative or resistive magnetospheres introduced in the literature. The energetic of the magnetosphere model will leave the degeneracy contrary to a pure geometrical study of light-curves. As phase-resolved spectra are available in radio and gamma-rays, we plan to compute multi-wavelength light-curves and spectra based on the above magnetosphere models.

The physics of radiative magnetospheres requires a more detailed investigation of the central role of particle injection and its crucial impacts on the magnetosphere energetics before to confront to the observations. Some more ingredients are required before investigating the gamma-ray light curves of individual pulsars.

A large amount of work has been carried out to identify these effects via in particle-in-cell simulations, taking the feedback between particle acceleration and radiation self-consistently into account. Relativistic magnetic reconnection is supposed to play a key role in the dissipation of the Poynting flux channelling into particle acceleration and emitting synchrotron photons as demonstrated by \cite{cerutti_modelling_2016} and \cite{philippov_ab-initio_2018}. The particle production assumption, either from the surface or from the light-cylinder or from the whole magnetosphere volume decides what on the global solution found in numerical simulations. \cite{chen_filling_2020} presented simulations not requiring efficient pair production in the vicinity of the light cylinder and found quasi-periodic solution able to explain the cone versus and core emission characteristics of radio pulses. Previously \cite{chen_electrodynamics_2014} already found significant curvature and synchrotron photon production around the current sheet of an aligned rotator. These results are able to reproduce the Fermi gamma-ray pulsar observations. In the same vain, \cite{kalapotharakos_fermi_2017} computed test particle trajectories and the associated radiation and particle efficiency in the so called FIDO model assuming curvature emission in the radiation reaction, in a similar way to the present study. Based on gamma-ray pulsar luminosity, they found a positive correlation between the pair multiplicity factor and the pulsar spin down.

Nevertheless almost all these simulations rely on particle injection prescriptions which are still not fully resolved from a physical point of view. As the magnetospheric solution tend to depends crucially on the injection rate, it is not yet clear how all these processes operate. Moreover, the Lorentz factors obtained are still many orders of magnitude below realistic values expected from observations. Our approach represents a good alternative to tackle this important issue of very high Lorentz factor within the magnetosphere.

\section*{Acknowledgements}

I am grateful to the referee for helpful comments and suggestions.
This work has been supported by the CEFIPRA grant IFC/F5904-B/2018 and ANR-20-CE31-0010. We acknowledge the High Performance Computing center of the University of Strasbourg for supporting this work by providing scientific support and access to computing resources.

\section*{Data availability}

The data underlying this article will be shared on reasonable request to the corresponding author.








%


\bsp	
\label{lastpage}
\end{document}